\documentstyle[epsfig]{mn2e}

\def\gtsima{$\; \buildrel > \over \sim \;$}
\def\ltsima{$\; \buildrel < \over \sim \;$}
\def\gtrsim{\lower.5ex\hbox{\gtsima}}
\def\lesssim{\lower.5ex\hbox{\ltsima}}

\begin{document}

\title[ULXs and remnants of massive metal-poor stars]{Ultra-luminous X-ray sources and remnants of massive metal-poor stars}
\author[Mapelli et al.]
{M. Mapelli$^{1,2}$, E. Ripamonti$^{2}$, L. Zampieri$^{3}$, M. Colpi$^{2}$, A. Bressan$^{3,4}$
\\
$^{1}$ Institute for Theoretical Physics, University of Z\"urich, Winterthurerstrasse 190, CH--8057, Z\"urich, Switzerland\\
$^2$Universit\`a Milano Bicocca, Dipartimento di Fisica G.Occhialini, Piazza
della Scienza 3, I--20126, Milano, Italy; {\tt mapelli@mib.infn.it}\\
$^3$INAF-Osservatorio astronomico di Padova, Vicolo dell'Osservatorio 5, I--35122, Padova, Italy\\
$^4$Scuola Internazionale Superiore di Studi Avanzati (SISSA), via Beirut 4,
I--34014, Trieste, Italy
}
\maketitle \vspace {7cm }

  \begin{abstract}
Massive metal-poor stars might form massive stellar black holes (BHs), with mass $25\le{}m_{\rm BH}/{\rm M}_{\odot{}}\le{}80$, via direct collapse. We derive the number of massive BHs (N$_{\rm BH}$) that are expected to form per galaxy through this mechanism. Such massive BHs might power most of the observed ultra-luminous X-ray sources (ULXs). We  select a sample of  64 galaxies with X-ray coverage, measurements of the star formation rate (SFR) and of the metallicity. We find that N$_{\rm BH}$ correlates with the number of observed ULXs per galaxy (N$_{\rm ULX}$) in this sample. We discuss the dependence of our model on the SFR and on the metallicity. The SFR is found to be crucial, consistently with previous studies. The metallicity plays a role in  our model, since a lower metallicity enhances the formation of massive BHs.  Consistently with our model, the data indicate that there might be an anticorrelation between N$_{\rm ULX}$, normalized to the SFR, and the metallicity. 
A larger and more homogeneous sample of metallicity measurements is required, in order to confirm our results.
\end{abstract}

%

\section{Introduction}
Ultra-luminous X-ray sources (ULXs, see Mushotzky 2004 for a review, and references therein) are defined as  non-nuclear point-like sources with isotropic X-ray luminosity $L_{\rm X}\gtrsim{}10^{39}$ erg s$^{-1}$. 
The mechanism that powers the ULXs is still unknown, although various scenarios have been proposed. ULXs could be associated with high-mass X-ray binaries (HMXBs) powered by stellar-mass black holes (BHs) with anisotropic X-ray emission (e.g. King et al. 2001) or with super-Eddington accretion rate/luminosity (e.g. Begelman 2002; King \&{} Pounds 2003; Socrates \&{} Davis 2006; Poutanen et al. 2007) or with a combination of the two mechanisms (e.g. King 2008). 
ULXs could also be associated with HMXBs powered by intermediate-mass BHs (IMBHs), i.e. BHs with mass $100\,{}{\rm M}_\odot{}\le{}m_{\rm BH}\le{}10^5\,{}{\rm M}_\odot{}$ (see van der Marel 2004 for a review). 
IMBHs with mass larger than $100\,{}{\rm M}_\odot{}$ may be required to explain the brightest ULXs (i.e. the $\lesssim{}4$ ULXs with $L_{\rm X}\gtrsim{}10^{41}$ erg s$^{-1}$), those ULXs showing quasi-periodic oscillations (M82 X-1, see Strohmayer \&{} Mushotzky 2003, and NGC 5408 X-1, see Strohmayer et al. 2007) and some of those that are surrounded by isotropically ionized nebulae (Pakull \&{} Mirioni 2002; Kaaret, Ward \&{} Zezas 2004). However, IMBHs are not needed to explain the observational properties of most of the ULXs (e.g. Gon\c{c}alves \&{} Soria 2006; Stobbart, Roberts \&{} Wilms 2006; Copperwheat et al. 2007; Roberts 2007;  Zampieri \&{} Roberts 2009). Thus, the objects classified as ULXs might actually be an inhomogeneous sample, including sources of different nature.

 Most of ULXs are located in galaxies with a high star formation rate (SFR, e.g.  Irwin, Bregman \&{} Athey  2004), although a small fraction (10--20 per cent, especially in the low-luminosity tail) might be associated with an old population (e.g. Colbert et al. 2004; Brassington, Read \&{} Ponman 2005). The ULXs match the correlation between X-ray luminosity and SFR reported by various studies (Grimm, Gilfanov \&{} Sunyaev 2003; Ranalli, Comastri \&{} Setti 2003; Gilfanov, Grimm \&{} Sunyaev 2004a,b,c; Kaaret \&{} Alonso-Herrero 2008). Furthermore, the same studies indicate that the luminosity function of ULXs is the direct extension of that of HMXBs. Recent papers suggest a correlation between ULXs and low-metallicity environments, and propose that this may be connected with the influence of metallicity on the evolution of massive stars (Pakull \&{} Mirioni 2002; Zampieri et al. 2004; Soria et al. 2005; Swartz, Soria \&{} Tennant 2008).
This scenario has been explored in detail by Mapelli, Colpi \&{} Zampieri (2009, hereafter Paper~I) and by Zampieri \&{} Roberts (2009), highlighting that a large fraction of ULXs may actually host massive
($\sim{}30-80\,{}{\rm M}_{\odot{}}$) stellar BHs formed in a low-metallicity environment.
In fact, low-metallicity massive ($\gtrsim{}40\,{}{\rm M}_\odot{}$) stars lose only a small fraction of their mass due to stellar winds (Maeder 1992, hereafter M92; Heger \&{} Woosley 2002, hereafter HW02; Heger et al. 2003, hereafter H03; Belczynski et al. 2010, hereafter B10) and can directly collapse (Fryer 1999; B10) into massive BHs ($25\,{}{\rm M}_\odot{}\le{}m_{\rm BH}\le{}80\,{}{\rm M}_\odot{}$). Such massive BHs can power  most of the known ULXs without requiring super-Eddington accretion or anisotropic emission. Furthermore, their formation mechanism can explain the correlation between ULXs and SFR, and the fact that ULXs are preferentially found in low-metallicity  regions. 
In this Paper, we extend to a larger sample of galaxies the analysis reported in Paper~I, and we study the formation of massive BHs from the direct collapse of massive metal-poor stars and their possible connection with ULXs. In particular, we show that there is a 
correlation between the number of massive BHs formed in this scenario and the number of observed ULXs per galaxy.


\section{Method}
The aim of this Paper is to compute the number of massive BHs that are expected to form in a galaxy (${\rm N}_{\rm BH}$) through the direct collapse of massive metal-poor stars. We will  then compare such number with the observed number of ULXs (${\rm N}_{\rm ULX}$) in the same galaxy and with other observational quantities (e.g. the SFR, the metallicity, etc.). In this Section, we start describing the procedure adopted for deriving ${\rm N}_{\rm BH}$ and then present the properties of the galaxy sample used for the comparison.
\subsection{The number of massive BHs}
According to numerical models (Fryer 1999; HW02; H03), a star that, at the end of its life, has a final mass $m_{\rm fin}\ge{}40\,{}{\rm M}_\odot{}$ is expected to directly collapse into a BH. In this case, the mass of the remnant BH is likely  more than half of
the final mass of the progenitor star, as 
  relatively small mass ejection is expected in the direct collapse. Thus, stars that at the end of their lives have $m_{\rm fin}\ge{}40\,{}{\rm M}_\odot{}$ are likely to produce massive BHs (B10).

The final masses of the stars strongly depend on their metallicity. Massive stars with metallicity close to solar cannot have final masses larger than $m_{\rm fin}\sim{}10-15\,{}{\rm M}_\odot{}$, even if their initial mass was very large, as they are expected to lose a lot of mass due to stellar winds (H03). Instead, massive stars with lower metallicity are less affected by stellar winds, and retain a larger fraction of their initial mass. If its metallicity is sufficiently low, a star can have a final mass $m_{\rm fin}\ge{}40\,{}{\rm M}_\odot{}$ and can directly collapse into a massive BH with a mass $25\,{}{\rm M}_\odot{}\le{}m_{\rm BH}\le{}80\,{}{\rm M}_\odot{}$ (HW02; H03; B10).

Thus, in order to compute the number of massive BHs ${\rm N}_{\rm BH}$, we first need to know for which metallicities massive stars can directly collapse into BHs. 
Evolutionary tracks of massive stars have been computed under different
assumptions concerning the mass-loss rate (M92; Bressan et al 1993; Fagotto et al 1994a,b; Eldridge \&{} Tout 2004; Eldridge \&{} Vink 2006; B10), and thus predict  different final stellar masses (differences are up to a factor of 2).
 In Paper~I, we used the results from M92, that can be regarded as upper limits. In this Paper, we base our calculations on the results of two previous studies: one relies on the PADOVA tracks (as described in Portinari, Chiosi \& Bressan 1998, hereafter referred to as P98) and the other is presented by B10. 
P98 assume that all massive stars explode as supernovae, and do not consider the direct collapse scenario.
 Here, we follow the more recent studies  by Fryer (1999) and by H03, indicating that,
above 40 M$_\odot$, the  stars collapse to a massive BH.
Thus, in the case of P98, we take the final stellar masses for various metallicities listed in their table~1, and we assume that all stars with $m_{\rm fin}\ge{}40\,{}{\rm M}_\odot{}$ end up into a massive BH\footnote{An important difference from the original assumptions made by P98 concerns the fate of very massive stars, with Hydrogen exhausted core at the end of the
hydrostatic evolution, M$_{\rm He}$, larger than 40 M$_\odot$.
For  M$_{\rm He}>$40 M$_\odot$, P98 adopted the pair instability (Woosley \&{} Weaver 1986) model with the following fates:
for  M$_{\rm He}$ in the range  40 to 60 M$_\odot$ the star undergoes a  pulsational instability leading to a final iron core collapse;
in the range M$_{\rm He}$ $\sim$60  to $\sim$100 M$_\odot$ the star was supposed to undergo a complete disruption and finally, for even larger cores, a  direct collapse  to a BH.}. 

Instead, B10 already include in their code the hypothesis that massive stars directly collapse into massive BHs. Thus, when we apply the results from B10 to our model, we consider the stars that end into BHs with $m_{\rm BH}\gtrsim{}25\,{}{\rm M}_\odot{}$ as progenitors of massive BHs (see fig.~1 of B10). This condition is approximately equivalent\footnote{We assume $m_{\rm BH}\gtrsim{}25\,{}{\rm M}_\odot{}$ for B10, because
 the models (e.g HW02; H03) show that in the direct collapse slightly more than half of the final mass of the star goes into the BH. As the minimum final stellar mass for direct collapse is $m_{\rm fin}\ge{}40\,{}{\rm M}_\odot{}$, $m_{\rm BH}\gtrsim{}20-30\,{}{\rm M}_\odot{}$ is a reasonable choice. 
On the basis of this argument, threshold masses anywhere in between 20 and 30 M$\odot{}$ would be fine. We choose $m_{\rm BH}\gtrsim{}25\,{}{\rm M}_\odot{}$ as a reasonable fiducial value. Furthermore, BHs with $m_{\rm BH}\gtrsim{}25\,{}{\rm M}_\odot{}$ are above the observed mass-range for stellar BHs ($3-20\,{}{\rm M}_\odot{}$, Orosz 2003).}  to requiring that $m_{\rm fin}\ge{}40\,{}{\rm M}_\odot{}$.

Given these assumptions, we can derive the expected number of massive BHs per galaxy (${\rm N}_{\rm BH}$) as a function of the star formation rate (SFR) and of the metallicity $Z$ (see Paper~I):
\begin{equation}\label{eq:totnum}
{\rm N}_{\rm BH}({\rm SFR},\,{}Z)=A({\rm SFR})\,{}\,{}\,{}\int_{m_{\rm prog}(Z)}^{m_{\rm max}}m^{-\alpha{}}\,{}{\rm d}m,
\end{equation}
where $m_{\rm max}$ is the maximum stellar mass (we assume $m_{\rm max}=120\,{}{\rm M}_\odot{}$) and  $\alpha{}$ is the  slope  of the initial mass function (IMF). 
$m_{\rm prog}(Z)$ is the minimum initial stellar mass (i.e. the mass at zero-age main sequence) for which a star is the progenitor of a massive BH. As we discussed above, $m_{\rm prog}(Z)$ strongly depends on the metallicity. In our calculations, we assume $m_{\rm prog}(Z)$ to be the initial stellar mass for which $m_{\rm fin}\ge{}40\,{}{\rm M}_\odot{}$ and  $m_{\rm BH}\gtrsim{}25\,{}{\rm M}_\odot{}$, when adopting the models by P98 and by B10, respectively.
 Finally, $A({\rm SFR})$, the normalization constant in equation~(\ref{eq:totnum}), can be estimated as
\begin{equation}\label{eq:norm1}
A({\rm SFR})=\frac{\int_{0}^{t_{\rm SFR}}{\rm SFR}(t)\,{}\,{}\,{}{\rm d}t}{\int_{m_{\rm min}}^{m_{\rm max}}m^{1-\alpha{}}\,{}{\rm d}m}, 
\end{equation}
where $m_{\rm min}$ is the minimum stellar mass (we assume $m_{\rm min}=0.08$ ${\rm M}_\odot{}$), SFR$(t)$ is the star formation rate as a function of time  and $t_{\rm SFR}$ is the duration of the star formation.

However, we are not interested in all the massive BHs, but only in those that could acquire massive stellar companions and power observable ULXs. Thus, in the case of interest, equation~(\ref{eq:norm1}) might be written as
\begin{equation}\label{eq:norm}
A({\rm SFR})=\frac{{\rm SFR}\,{}\,{}\,{}t_{\rm co}}{\int_{m_{\rm min}}^{m_{\rm max}}m^{1-\alpha{}}\,{}{\rm d}m}, 
\end{equation}
where SFR is the current star formation rate and $t_{\rm co}$ is the characteristic lifetime of the companion of the massive BH. 
In this Paper, we adopt a constant value $t_{\rm co}=10^7$ yr. Such value is the lifetime of a $\sim{}15\,{}{\rm M}_\odot$ star. In fact, according to theoretical models, massive BHs can power persistent ULXs only when their companion is sufficiently massive ($\ge{}15\,{}{\rm M}_\odot{}$, Patruno et al. 2005). If the companion star is relatively low-mass ($2\le{}m/{\rm M}_\odot{}<10$), then the system can power only a transient ULX, with an outburst phase of a few days every few months\footnote{However, the Galactic transient source GRS~$1915+105$, which is associated to a low-mass X-ray binary system including a quite massive BH (m$_{\rm BH}=14.0\pm{}4.4$ M$_{\odot{}}$, Harlaftis \&{} Greiner 2004), has a much longer outburst phase ($>17$ years, see e.g. Castro-Tirado, Brandt \&{} Lund 1992; Fender \&{} Belloni 2004; Deegan, Combet \&{} Wynn 2009), in partial contradiction with this model.} (Portegies Zwart, Dewi \&{} Maccarone 2004). The probability of observing such transient ULXs is a factor of $\gtrsim{}10^2$ lower than the probability of observing persistent ULXs. 
Thus, the value of ${\rm N}_{\rm BH}$ derived assuming $t_{\rm co}=10^7$ yr is not the total number of massive BHs present in a galaxy, but the number of massive BHs that could easily find a massive companion to accrete from.

\begin{figure}
\center{{
\epsfig{figure=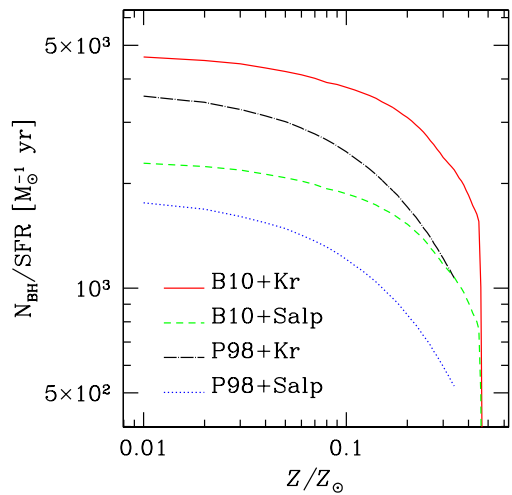,height=7cm} 
}}
\caption{\label{fig:fig1}
Number of expected massive BHs per galaxy, normalized to the SFR, as a function of $Z$. Solid line (red on the web): B10 with the Kroupa IMF; dashed line (green on the web): B10 with the Salpeter IMF; dot-dashed line (black): P98 with the Kroupa IMF; dotted line (blue on the web): P98 with the Salpeter IMF.}
\end{figure}

 Fig.~\ref{fig:fig1} shows ${\rm N}_{\rm BH}$, normalized to the SFR, as a function of $Z$, for P98 and for B10. In this Fig., we adopt two different IMFs: (i) the Salpeter IMF (Salpeter 1955), for which $\alpha{}=2.35$, and (ii) the Kroupa IMF, for which $\alpha{}=1.3$ if $m\le0.5\,{}{\rm M}_\odot{}$ and $\alpha{}=2.3$ for larger masses (Kroupa 2001). Fig.~\ref{fig:fig1} indicates that ${\rm N}_{\rm BH}$ has approximately the same trend when adopting the Salpeter or the Kroupa IMF, although the normalization is different by a factor of $\sim{}2$. 
In the following, we will consider, as reference, only the Kroupa IMF. As shown in Appendix A, where a more detailed comparison is reported, the results can be easily scaled to the Salpeter IMF.
Fig.~\ref{fig:fig1} shows also that the trend obtained adopting P98 is quite similar to the one obtained adopting B10. The main differences are in the normalization, as ${\rm N}_{\rm BH}$ is a factor of $\sim{}1.3-2.0$ lower for P98 than for B10, and in the cut-off, as ${\rm N}_{\rm BH}$  goes to zero for $Z\sim{}0.4\, Z_\odot$ and $0.5\, Z_\odot$ for P98 and B10, respectively.

\begin{table*}
\begin{center}
\caption{Properties of the galaxies in our sample.} \leavevmode
\begin{tabular}[!h]{lllllll}
\hline
Galaxy
& SFR (${\rm M}_\odot{}$ yr$^{-1}$)
& $Z$ ($Z_\odot{}$)
& N$_{{\rm ULX,\,{}raw}}$ $^{\rm a}$
& D (Mpc) $^{\rm b}$
& $Q$ $^{\rm c}$
& N$_{\rm ULX}$ $^{\rm d}$\\
\hline
The Cartwheel       & 20                          & 0.14           & 19  & 124   & 0 & $19.00_{-4.32}^{+5.43}$   \\
NGC~253             & 4.0                         & 0.24           & 3   & 3.0   & $0.02_{-0.01}^{+0.04}$ & $2.98_{-1.63}^{+2.91}$   \\
NGC~300	            & 0.14	                  & 0.19	   & 0   & 2.02  & $0$ & $0.00_{-0.00}^{+1.83}$ \\
NGC~598 (M~33) 	    & 1.1                         & 0.32           & 1   & 0.92  & $0.00_{-0.00}^{+0.00}$ & $1.00_{-0.83}^{+2.29}$   \\
NGC~628 (M~74)       & 2.2                        & 0.27           & 2   & 8.5   & $0.15_{-0.06}^{+0.11}$ & $1.85_{-1.29}^{+2.63}$    \\
NGC~1058	    & 0.27                        & 0.26           & 1   & 9.85  & $0.09_{-0.03}^{+0.06}$ & $0.91_{-0.83}^{+2.29}$ \\
NGC~1073            & 1.2                         & 0.34           & 2   & 15.2  & 0 &  $2.00_{-1.29}^{+2.63}$  \\
NGC~1291            & 0.94                        & 0.06           & 3   & 8.6   & $0.16_{-0.07}^{+0.11}$ & $2.84_{-1.63}^{+2.91}$ \\
NGC~1313            & 1.4                         & 0.1            & 2   & 3.7   & 0 &  $2.00_{-1.29}^{+2.63}$     \\
NGC~1365	    & 7.2                         & 0.20           & 13  & 17.4  & $0.82_{-0.16}^{+0.20}$ & $12.18_{-3.56}^{+4.70}$   \\
IC~342              & 0.48                        & 0.19           & 2   & 3.9   & $0.02_{-0.01}^{+0.04}$ & $1.98_{-1.29}^{+2.63}$   \\
NGC~1566	    & 3.2                         & 0.33           & 4   & 13.4  & $1.50_{-0.13}^{+0.18}$ & $2.50_{-1.92}^{+3.16}$\\
NGC~1705	    & 0.087	                  & 0.19           & 0   & 6.0   & $0$ & $0.00_{-0.00}^{+1.83}$   \\
NGC~2366            & 0.075              	  & 0.10           & 0	 & 2.7   & $0$ & $0.00_{-0.00}^{+1.83}$    \\
NGC~2403	    & 0.4                         & 0.22           & 1   & 3.1   & 0 & $1.00_{-0.83}^{+2.29}$   \\
NGC~2442            & 4.6                         & 0.45           & 2   & 17.1  & $1.33_{-0.31}^{+0.10}$ & $0.67_{-0.67}^{+2.65}$\\
Holmberg~II (Arp~268)  & 0.1                      & 0.1            & 1   & 4.5   & 0 & $1.00_{-0.83}^{+2.29}$   \\
NGC 2903            & 1.9	                  & 0.28           & 2	 & 7.4   & $0.09_{-0.04}^{+0.07}$ & $1.91_{-1.29}^{+2.63}$\\
NGC~3031 (M~81)	    & 3.3                         & 0.29           & 2   & 3.4   & $0.03_{-0.02}^{+0.05}$ & $1.97_{-1.29}^{+2.63}$  \\ 
NGC~3049            & 0.57                        & 0.60           & 0   & 19.9  & $0$ & $0.00_{-0.00}^{+1.83}$\\
PGC~30819 (IC~2574, UGC~05666) & 0.057            & 0.17           & 0   & 3.5   & $0$ & $0.00_{-0.00}^{+1.83}$\\  
NGC~3310 (Arp~217)  & 2.2                         & 0.22           & 3   & 18.7  & $0.16_{-0.03}^{+0.03}$ & $2.84_{-1.63}^{+2.91}$\\ 
NGC~3395/3396 (Arp~270) & 4.7                     & 0.21           & 7   & 24.6  & 0 & $7.00_{-2.58}^{+3.76}$\\
PGC~35286 (UGC~06456) & 0.017                     & 0.062          & 0   & 1.4   & $0$ & $0.00_{-0.00}^{+1.83}$\\
PGC~35684 (UGC~06541, Mkn~178) & 0.01             & 0.091          & 0   & 4.0   & $0$ & $0.00_{-0.00}^{+1.83}$\\
NGC~3738 (Arp~234)  & 0.038                       & 0.20	   & 0	 & 4.3   & $0$ & $0.00_{-0.00}^{+1.83}$\\
NGC~3972  	    & 0.30                        & 0.35           & 0	 & 17.0  & $0$ & $0.00_{-0.00}^{+1.83}$\\
Antennae (NGC~4038/4039, Arp~244)  &  7.1         & 0.04           & 15  & 21.5  & 0 & $15.00_{-3.83}^{+4.95}$\\
NGC~4144            & 0.05                        & 0.20           & 0   & 5.7   & $0$ & $0.00_{-0.00}^{+1.83}$\\
NGC~4214            & 0.13                        & 0.21           & 0   & 3.5   & $0$ & $0.00_{-0.00}^{+1.83}$\\
NGC~4236            & 0.12                        & 0.22           & 0   & 2.2   & $0$ & $0.00_{-0.00}^{+1.83}$\\
NGC~4248            & 0.018                       & 0.26           & 0   & 7.3   & $0$ & $0.00_{-0.00}^{+1.83}$\\
NGC~4254 (M~99)	    & 4.0                         & 0.26           & 3   & 16.8  & $1.39_{-0.13}^{+0.11}$ & $1.61_{-1.61}^{+2.91}$\\
NGC~4258 (M~106)	    & 2.5                 & 0.32           & 3   & 7.7   & $0.58_{-0.25}^{+0.45}$ & $2.42_{-1.69}^{+2.92}$\\
NGC~4303 (M~61)	    & 5.8                         & 0.20           & 6   & 15.2  & $1.91_{-0.45}^{+0.58}$ & $4.09_{-2.45}^{+3.60}$\\
NGC~4321 (M~100)     & 4.8	                  & 0.28           & 7   & 14.1  & $1.88_{-0.49}^{+0.65}$ & $5.12_{-2.66}^{+3.79}$\\
NGC~4395	    & 0.12                        & 0.22           & 1   & 3.6   & $0.04_{-0.03}^{+0.07}$ & $0.96_{-0.83}^{+2.29}$\\
NGC~4449            & 0.28                        & 0.25           & 0   & 3.0   & $0$ & $0.00_{-0.00}^{+1.83}$\\
NGC~4485/4490 (Arp~269) & 4.5                     & 0.23           & 5   & 8.55  & $0.16_{-0.06}^{+0.11}$ & $4.84_{-2.16}^{+3.38}$\\
NGC 4501 (M~88)$^{\rm e}$	       & 3.5      & 0.49           & 2   & 16.8  & $0.43_{-0.09}^{+0.11}$ & $1.57_{-1.29}^{+2.63}$\\
NGC~4559            & 1.3                         & 0.19           & 2   & 5.8   & 0 & $2.00_{-1.29}^{+2.63}$\\
NGC~4631 (Arp~281)  & 2.80                        & 0.22           & 2   & 6.9   & $0.16_{-0.08}^{+0.14}$ & $1.84_{-1.30}^{+2.63}$\\ 
NGC~4651 (Arp~189)  & 1.40                        & 0.23           & 1   & 16.8  & $0.17_{-0.03}^{+0.04}$ & $0.83_{-0.83}^{+2.29}$\\
NGC~4656            & 0.95                        & 0.075          & 1   & 7.2   & $0.05_{-0.02}^{+0.04}$ & $0.95_{-0.83}^{+2.29}$\\
The Mice (NGC~4676, Arp~242)  & 6.0               & 0.3            & 6   & 90.4 & 0 & $6.00_{-2.37}^{+3.58}$\\
NGC~4736 (M~94)      & 1.1                        & 0.20           & 1   & 5.2   & $0.15_{-0.08}^{+0.17}$ & $0.85_{-0.85}^{+2.29}$\\
NGC~4861 (Arp~266)  & 0.59                        & 0.13           & 2   & 17.8  & $0.11_{-0.02}^{+0.03}$ & $1.89_{-1.29}^{+2.63}$\\
PGC~45561 (UGC~08231) & 0.29                      & 0.40           & 0   & 33.0  & $0$ & $0.00_{-0.00}^{+1.83}$\\
NGC~5033	    & 1.50                        & 0.069          & 2   & 15.2  & $0.69_{-0.16}^{+0.21}$ & $1.31_{-1.30}^{+2.63}$\\
NGC~5055 (M~63)	    & 1.70                        & 0.30           & 2   & 8.5   & $0.18_{-0.07}^{+0.13}$ & $1.82_{-1.29}^{+2.63}$\\
NGC~5194/5195 (M~51, Arp~85) & 13.0               & 0.39           & 9   & 7.7   & $0.11_{-0.05}^{+0.08}$ & $8.89_{-2.94}^{+4.10}$\\
NGC~5236 (M~83)      & 2.6                        & 0.36           & 1   & 4.7   & $0.09_{-0.05}^{+0.12}$ & $0.91_{-0.84}^{+2.29}$\\
NGC~5238 (Mkn~1479) & 0.013                       & 0.26           & 0   & 4.9   & $0$ & $0.00_{-0.00}^{+1.83}$\\
NGC~5408            & 0.09                        & 0.11           & 1   & 4.85  & 0 & $1.00_{-0.83}^{+2.29}$\\
NGC~5457 (M~101, Arp~26) & 3.1                    & 0.17           & 7   & 6.9   & $0.40_{-0.18}^{+0.34}$ & $6.60_{-2.60}^{+3.77}$\\
\noalign{\vspace{0.1cm}}
\hline
\end{tabular}
\footnotesize{\\{\it Table~1 - continued.}}
\end{center}
\end{table*}
\begin{table*}
\begin{center}
\footnotesize{{\it Table~1 - continued.}\\}
\begin{tabular}[!h]{lllllll}
\hline
Galaxy
& SFR (${\rm M}_\odot{}$ yr$^{-1}$)
& $Z$ ($Z_\odot{}$)
& N$_{{\rm ULX,\,{}raw}}$ $^{\rm a}$
& D (Mpc) $^{\rm b}$
& $Q$ $^{\rm c}$
& N$_{\rm ULX}$ $^{\rm d}$\\
\hline
Circinus	    & 1.5                         & 0.1            & 4   & 4.2   & $0.00_{-0.00}^{+0.01}$ & $4.00_{-1.91}^{+3.15}$\\
NGC~6946 (Arp~29)   & 3.56                        & 0.283          & 3   & 5.5   & $0.08_{-0.05}^{+0.11}$ & $2.92_{-1.63}^{+2.91}$\\
PGC~68618 (IC~5201) & 1.7                         & 0.260          & 1   & 11.1  & $0.49_{-0.16}^{+0.24}$ & $0.51_{-0.51}^{+2.29}$\\
NGC~7714/7715 (Arp~284, Mkn~538) & 7.2            & 0.2            & 9   & 36.6  & $0.84_{-0.12}^{+0.14}$ & $8.16_{-2.94}^{+4.11}$\\%
NGC~7742	    & 1.27	                  & 0.245          & 2   & 22.2  & $0.32_{-0.05}^{+0.05}$ & $1.68_{-1.29}^{+2.63}$\\
Milky Way	    & 0.25                        & 0.306          & 0   & $-$   & 0 & $0.00_{-0.00}^{+1.83}$\\
IC~10		    & $7.14\times{}10^{-2}$       & 0.22           & 0   & 0.70  & 0 & $0.00_{-0.00}^{+1.83}$\\
Large Magellanic Cloud (LMC) & 0.25               & 0.27           & 0   & 0.051 & 0 & $0.00_{-0.00}^{+1.83}$\\
Small Magellanic Cloud (SMC) & 0.15               & 0.129          & 0   & 0.064 & 0 & $0.00_{-0.00}^{+1.83}$\\
\noalign{\vspace{0.1cm}}
\hline
\end{tabular}
\end{center}
\footnotesize{{\it Notes.} The references for the values of SFR, $Z$ and N$_{{\rm ULX,\,{}raw}}$ reported in this Table are given in Appendix~B. $^{\rm a}$ N$_{{\rm ULX,\,{}raw}}$ is the observed number of ULXs per galaxy, before the subtraction of contaminating sources. $^{\rm b}$ D is the distance in Mpc (from the Liu \& Bregman 2005 catalogue; if the galaxy is not in such catalogue, from the NASA Extragalactic Database).
$^{\rm c}$ Number of expected contaminating objects, estimated as described in Sections 2.2.1 and B4 (Appendix B). When $Q$ is equal to zero (without uncertainties), either the contamination was already subtracted in N$_{{\rm ULX,\,{}raw}}$ (i.e. in the paper with the original data) and N$_{{\rm ULX,\,{}raw}}={\rm N}_{\rm ULX}$, or N$_{{\rm ULX,\,{}raw}}=0$. $^{\rm d}$ N$_{\rm ULX}$ is the observed number of ULXs per galaxy, after the subtraction of the estimated number of contaminating sources. The errors come from the combination of the Poissonian uncertainties upon N$_{{\rm ULX,\,{}raw}}$ (according to the treatment in Gehrels 1986), and on $Q$ (with uncertainties from the Hasinger et al. 1998 $\log({\rm N})-\log({\rm S})$ relation).
$^{\rm e}$ NGC~4501 is the only galaxy with N$_{\rm ULX}>0$ and with $Z>0.47\,{}Z_\odot{}$. For this galaxy N$_{\rm BH}=0$ in both P98 and B10 models (see Fig.~\ref{fig:fig1} and Section~3.2), although  N$_{\rm BH}+1\,{}\sigma{}>0$.}
\end{table*}
\begin{table*}
\begin{center}
\caption{Values of N$_{\rm BH}$ and $\epsilon_{\rm BH}$ for the galaxies listed in Table~1, assuming the models by P98 (first and second column) and by B10 (third and fourth column). We adopt a Kroupa IMF.} \leavevmode
\begin{tabular}[!h]{lllll}
\hline
Galaxy
& N$_{\rm BH}$ (P98)
& $\epsilon_{\rm BH}/10^{-4}$ (P98)
& N$_{\rm BH}$ (B10)
& $\epsilon_{\rm BH}/10^{-4}$ (B10)\\
\hline
The Cartwheel       & 41900$^{+21700}_{-21600}$      & $4.5^{+2.7}_{-2.6}$     &  70200$^{+35200}_{-35400}$  &  2.7$^{+1.6}_{-1.5}$\\
NGC~253             & 6070$^{+3180}_{-3240}$         & $4.9^{+5.4}_{-3.8}$     &  11200$^{+5700}_{-5900}$    &  $2.7^{+2.9}_{-2.0}$ \\
NGC~300             & 245$^{+125}_{-130}$            & $0^{+74.7}_{-0}$        &  438$^{+222}_{-224}$        &  $0^{+41.8}_{-0}$ \\
NGC~598       	    & 1280$^{+710}_{-1280}$          & $7.8^{+18.4}_{-7.8}$    &  2510$^{+1330}_{-1380}$     &  $4.0^{+9.4}_{-4.0}$\\
NGC~628             & 2860$^{+1570}_{-1550}$         & $6.7^{+9.9}_{-5.9}$     &  5440$^{+2880}_{-2850}$     &  $3.5^{+5.2}_{-3.0}$\\
NGC~1058	    & 377$^{+198}_{-203}$            & $24.1^{+62.1}_{-24.1}$  &  709$^{+374}_{-371}$        &  $12.8^{+33.0}_{-12.8}$ \\
NGC~1073            & 1290$^{+730}_{-1290}$          & $15.5^{+22.2}_{-15.5}$    &  2620$^{+1390}_{-1450}$   &  $7.6^{+10.8}_{-6.5}$ \\
NGC~1291	    & 2730$^{+1370}_{-1380}$         & $10.6^{+11.9}_{-8.1}$   &  3690$^{+1860}_{-1850}$     &  $7.9^{+8.8}_{-6.0}$\\
NGC~1313            & 3460$^{+1760}_{-1750}$         & $5.8^{+8.1}_{-4.7}$     &  5100$^{+2580}_{-2560}$     &  $3.9^{+5.5}_{-3.2}$\\
NGC~1365	    & 12600$^{+6400}_{-6700}$        & $6.5^{+4.7}_{-4.2}$    &  20900$^{+11000}_{-10600}$     &  $5.8^{+3.8}_{-3.4}$\\
IC~342              & 841$^{+451}_{-435}$            & $23.5^{+33.7}_{-19.7}$  &  1500$^{+760}_{-770}$       &  $13.2^{+18.7}_{-11.0}$\\
NGC~1566            & 3560$^{+1990}_{-3560}$         & $7.0^{+10.3}_{-7.0}$     &  6780$^{+3730}_{-3570}$    &  $3.7^{+5.4}_{-3.7}$ \\
NGC~1705            & 153$^{+78}_{-81}$              & $0^{+119.6}_{-0}$       &  274$^{+139}_{-142}$        &  $0^{+66.8}_{-0}$  \\
NGC~2366            & 185$^{+94}_{-95}$              & $0^{+98.9}_{-0}$        &  272$^{+138}_{-136}$        &  $0^{+67.3}_{-0}$  \\
NGC~2403	    & 637$^{+333}_{-339}$            & $15.7^{+36.9}_{-15.5}$  &  1160$^{+600}_{-600}$       &  $8.6^{+20.2}_{-8.4}$\\
NGC~2442            & 0$^{+5270}_{-0}$               & $>1.3$                  &  5930$^{+4370}_{-5930}$     &  $1.1^{+4.9}_{-1.1}$\\
Holmberg~II         & 247$^{+125}_{-125}$            & $40.4^{+95.00}_{-39.4}$ &  364$^{+184}_{-182}$        &  $27.5^{+64.4}_{-26.6}$\\
NGC~2903            & 2440$^{+1340}_{-1370}$         & $7.8^{+11.6}_{-7.0}$     &  4640$^{+2450}_{-2430}$     &  $4.1^{+6.1}_{-3.6}$ \\
NGC~3031            & 4300$^{+2360}_{-2320}$         & $4.6^{+6.6}_{-3.9}$     &  8180$^{+4330}_{-4280}$     &  $2.4^{+3.5}_{-2.0}$ \\
NGC~3049            & 0$^{+0}_{-0}$                  & $-$                     &  0$^{+0}_{-0}$              &  $-$ \\
PGC~30819           & 110$^{+56}_{-58}$              & $0^{+166.4}_{-0}$       &  187$^{+95}_{-96}$          &  $0^{+97.9}_{-0}$\\
NGC~3310            & 3510$^{+1890}_{-1870}$         & $8.1^{+9.4}_{-6.4}$     & 6380$^{+3280}_{-3320}$      &  $4.4^{+5.1}_{-3.5}$ \\
NGC~3395/3396       & 7490$^{+4040}_{-3990}$         & $9.3^{+7.1}_{-6.1}$     & 13600$^{+7000}_{-7000}$     &  $5.1^{+3.8}_{-3.2}$ \\
PGC~35286           & 48$^{+24}_{-24}$               & $0^{+381.2}_{-0}$       & 68$^{+34}_{-34}$            &  $0^{+269.1}_{-0}$\\
PGC~35684           & 25$^{+13}_{-13}$               & $0^{+732.0}_{-0}$       & 38$^{+19}_{-19}$            &  $0^{+481.6}_{-0}$\\
NGC~3738            & 63$^{+33}_{-33}$               & $0^{+290.5}_{-0}$       & 114$^{+58}_{-58}$           &  $0^{+160.5}_{-0}$\\
NGC~3972            & 310$^{+174}_{-310}$            & $0^{+59.03}_{-0}$       & 628$^{+334}_{-348}$         &  $0^{+29.1}_{-0}$\\
Antennae            & 22200$^{+11100}_{-11200}$      & $6.7^{+4.0}_{-3.8}$     & 30000$^{+15000}_{-15000}$   &  $5.0^{+3.0}_{-2.8}$\\
NGC~4144            & 84$^{+44}_{-45}$               & $0^{+217.9}_{-0}$       & 151$^{+76}_{-77}$           &  $0^{+121.2}_{-0}$\\
NGC~4214            & 213$^{+115}_{-110}$            & $0^{+85.9}_{-0}$        & 404$^{+204}_{-210}$         &  $0^{+45.3}_{-0}$\\
NGC~4236            & 191$^{+100}_{-102}$            & $0^{+95.8}_{-0}$        & 348$^{+176}_{-181}$         &  $0^{+52.6}_{-0}$\\
NGC~4248            & 25 $^{+14}_{-14}$              & $0^{+732.0}_{-0}$       &  48$^{+25}_{-25}$           &  $0^{+381.2}_{-0}$\\
NGC~4254      	    & 5480$^{+2880}_{-2950}$         & $2.9^{+5.9}_{-2.9}$     &  10700$^{+5500}_{-5600}$    &  $1.5^{+3.0}_{-1.5}$\\
NGC~4258       	    & 2900$^{+1610}_{-2900}$         & $8.3^{+11.2}_{-8.3}$    &  5690$^{+3010}_{-3130}$     &  $4.3^{+5.7}_{-3.9}$\\
NGC~4303      	    & 9700$^{+5050}_{-5150}$         & $4.2^{+4.5}_{-3.6}$     &  16800$^{+8900}_{-8500}$    &  $2.4^{+2.6}_{-2.0}$\\
NGC~4321            & 6240$^{+3420}_{-3370}$         & $8.2^{+7.8}_{-6.4}$     &  11900$^{+6300}_{-6200}$    &  $4.3^{+4.1}_{-3.3}$\\
NGC~4395	    & 197$^{+107}_{-105}$	     & $48.7^{+119.2}_{-48.7}$ &  359$^{+185}_{-187}$        &  $26.7^{+65.3}_{-27.5}$\\
NGC~4449            & 408$^{+214}_{-219}$            & $0^{+44.8}_{-0}$        &  759$^{+400}_{-396}$        &  $0^{+24.1}_{-0}$ \\
NGC~4485/4490       & 6870$^{+3600}_{-3670}$         & $7.1^{+6.2}_{-5.0}$     &  12600$^{+6500}_{-6600}$    &  $3.9^{+3.3}_{-2.7}$ \\
NGC~4501            & 0$^{+3180}_{-0}$               & $>4.1$                  &  0$^{+6710}_{-0}$           &  $>1.9$ \\
NGC~4559            & 2370$^{+1230}_{-1260}$         & $8.4^{+11.9}_{-7.0}$    &  4240$^{+2150}_{-2200}$     &  $4.7^{+6.6}_{-3.9}$\\
NGC~4631            & 4450$^{+2400}_{-2370}$         & $4.1^{+6.3}_{-3.7}$     &  8090$^{+4160}_{-4220}$     &  $2.3^{+3.5}_{-2.0}$ \\
NGC~4651            & 2080$^{+1130}_{-1080}$         & $4.0^{+11.2}_{-4.0}$    &  3970$^{+2010}_{-2070}$     &  $2.1^{+5.9}_{-2.1}$\\
NGC~4656            & 2550$^{+1290}_{-1290}$         & $3.7^{+9.2}_{-3.7}$    &  3720$^{+1870}_{-1870}$     &  $2.6^{+6.3}_{-2.6}$\\
The Mice            & 6980$^{+3870}_{-6980}$         & $8.6^{+7.0}_{-8.6}$     &  13700$^{+7500}_{-7200}$    &  $4.4^{3.6}_{-2.9}$\\
NGC~4736            & 1930$^{+980}_{-1020}$          & $4.4^{+12.1}_{-4.4}$    &  3310$^{+1710}_{-1700}$     &  $2.6^{+7.0}_{-2.6}$\\
NGC~4861            & 1300$^{+670}_{-670}$           & $14.6^{+21.6}_{-12.7}$  &  2080$^{+1040}_{-1040}$     &  $9.1^{+13.4}_{-7.8}$ \\
PGC~45561           &    0$^{+377}_{-0}$             & $0^{+48.5}_{-0}$        &  531$^{+295}_{-531}$        &  $0^{+34.5}_{-0}$\\
NGC~5033	    & 4050$^{+2050}_{-2030}$         & $3.4^{+6.9}_{-3.4}$     &  5920$^{+2970}_{-2970}$     &  $2.3^{+4.7}_{-2.3}$\\
NGC~5055      	    & 2040$^{+1130}_{-2040}$         & $8.9^{+13.8}_{-8.9}$    &  3990$^{+2110}_{-2090}$     &  $4.6^{+7.0}_{-4.2}$\\
NGC~5194/5195       &    0$^{+17300}_{-0}$           & $>5.2$                  &  23300$^{+13600}_{-23300}$  &  $3.8^{+2.8}_{-3.8}$\\
NGC~5236            &    0$^{+3580}_{-0}$            & $>2.5$                  &  4990$^{+2660}_{-3620}$     &  $1.8^{+4.7}_{-1.8}$\\
NGC~5238            &   18$^{+10}_{-10}$             & $0{+1017.0}_{-0}$       &  35$^{+18}_{-18}$           &  $0^{+522.9}_{-0}$\\
NGC~5408            & 223$^{+113}_{-113}$            & $44.8^{+105.2}_{-43.6}$ &  328$^{+166}_{-164}$        &  $30.5^{+71.5}_{-29.6}$\\
NGC~5457       	    & 5890$^{+2990}_{-3110}$        & $11.2^{+8.6}_{-7.4}$     &  9610$^{+4870}_{-4860}$     &  $6.9^{+5.2}_{-4.4}$\\
\noalign{\vspace{0.1cm}}
\hline
\end{tabular}
\footnotesize{\\{\it Table~2 - continued.}}
\end{center}
\end{table*}
\begin{table*}
\begin{center}
\footnotesize{{\it Table~2 - continued.}\\}
\begin{tabular}[!h]{lllll}
\hline
Galaxy
& N$_{\rm BH}$ (P98)
& $\epsilon_{\rm BH}/10^{-4}$ (P98)
& N$_{\rm BH}$ (B10)
& $\epsilon_{\rm BH}/10^{-4}$ (B10)\\
\hline
Circinus	    & 3710$^{+1880}_{-1880}$	     & $10.8^{+10.1}_{-7.5}$   &  5460$^{+2760}_{-2740}$     &  $7.3^{+6.8}_{-5.1}$\\
NGC~6946	    & 4630$^{+2540}_{-2500}$       & $6.3^{+7.2}_{-4.9}$    &  $8820^{+4660}_{-4610}$     & $3.3^{+3.7}_{-2.5}$\\
PGC~68618           & 2330$^{+1220}_{-1250}$       & $2.2^{+10.1}_{-2.2}$   &  $4560^{+2350}_{-2430}$     & $1.1^{+5.2}_{-1.1}$\\
NGC~7714/7715       & 12600$^{+6400}_{-6700}$      & $9.7^{+6.2}_{-5.9}$    &  $22500^{+11400}_{-11700}$  & $3.6^{+2.6}_{-2.3}$\\
NGC~7742            & 1830$^{+1000}_{-980}$         & $9.2^{+15.3}_{-9.1}$   &  $3400^{+1800}_{-1780}$     & $4.9^{+8.2}_{-4.9}$\\
Milky Way	    & 291$^{+161}_{-291}$          & $0^{+62.9}_{-0}$       &  $570^{+302}_{-299}$        & $0^{+32.1}_{-0}$\\
IC~10		    & 114$^{+61}_{-61}$            & $0^{+160.5}_{-0}$      &  $207^{+107}_{-108}$        & $0^{+88.4}_{-0}$\\
LMC                 & 325$^{+178}_{-176}$          & $0^{+56.3}_{-0}$       &  $618^{+327}_{-323}$        & $0^{+29.6}_{-0}$  \\
SMC                 & 314$^{+167}_{-159}$          & $0^{+58.3}_{-0}$       &  $526^{+264}_{-264}$        & $0^{+34.8}_{-0}$\\  
\noalign{\vspace{0.1cm}}
\hline
\end{tabular}
\end{center}
\end{table*}

Once ${\rm N}_{\rm BH}$ is known, we can estimate the upper limit ($\epsilon{}_{\rm BH}$) of the fraction of massive BHs that power ULXs in a given galaxy at present, assuming that  all the observed ULXs in this galaxy, corrected for the contamination (${\rm N}_{\rm ULX}$, see next Section), are powered by a massive BH: 
\begin{equation}\label{eq:eobs}
\epsilon{}_{\rm BH}=\frac{{\rm N}_{\rm ULX}}{{\rm N}_{\rm BH}}.
\end{equation}

\subsection{The sample of galaxies}
In order to compare the value of ${\rm N}_{\rm BH}$ computed from equation~(\ref{eq:totnum}) with the observed number of ULXs per galaxy (${\rm N}_{\rm ULX}$),  we selected a sample of galaxies (Table~1 and references in Appendix~B) that satisfy the following criteria. (i) There exist X-ray observation(s) deep enough to detect the presence of ULXs. 
(ii) The selected galaxies have at least one measurement of the SFR. 
(iii) They have a sufficiently accurate measurement of the metallicity $Z$ (in most cases even of the metallicity gradient).
(iv) They are not elliptical galaxies.

The last criterion indicates that the galaxies in Table~1 are spiral, irregular or peculiar galaxies, but not elliptical galaxies. We decided not to include any elliptical galaxy in our sample, as ULXs in elliptical galaxies show properties that are quite different from those of ULXs in other galaxies, which may indicate a different origin for them (Swartz et al. 2004; Liu, Bregman \&{} Irwin 2006; Winter, Mushotzky \&{} Reynolds 2006). Furthermore, we remind that in early-type galaxies there is neither evidence of the presence of core-collapse supernovae nor of a significant fraction of massive star population, apart from small rejuvenation episodes (e.g. NGC~205, Bertola et al. 1995).  ULXs in elliptical galaxies might be connected to the minor (but non-negligible) fraction of ULXs that is claimed (e.g. Colbert et~al. 2004; Brassington et~al. 2005) to be associated with old stellar populations (hereafter, we refer to them as Population II ULXs). Since old populations are present also in spiral galaxies, our sample may be contaminated by Population II ULXs.  It is quite difficult to estimate their number, but it is reasonable to neglect their contribution in our sample (see Appendix~B for details).

 No complete catalogue of galaxies with and without ULXs exists. The most known available catalogues are Liu \&{} Mirabel (2005; hereafter LM05) and Liu \&{} Bregman (2005; hereafter LB05). LM05 is a catalogue of ULXs reported in the literature until April 2004. It does not include galaxies without ULXs, and it is quite non-uniform, as it reports observations with various instruments. LB05 is a catalogue of all the 313 nearby galaxies that have been observed with the {\it ROSAT} High Resolution Imager (HRI) with detection limits that would allow the detection of ULXs. Thus, the LB05 catalogue is less biased than LM05, as it includes  galaxies with and without ULXs, and it is more homogeneous, as all the data reported come from the same instrument. On the other hand, even LB05 suffers from some  observation bias  (i.e., galaxies were originally chosen to be observed for a myriad different reasons), and the instrumental properties (especially sensitivity and resolution) of {\it ROSAT} are quite worse than those of {\it Chandra} and {\it XMM-Newton}, whose observations are included in LM05. Furthermore, some galaxies have been observed after 2004 and are not included in these catalogues, but there is no reason not to take them into account.

 Given this situation, our sample is inevitably affected by various biases (e.g. galaxies without ULXs are likely under-represented) and it is not homogeneous. It is mainly based upon the LB05 catalogue, which includes 52 of the 64 galaxies listed in Table~1. The remaining 12 galaxies can be divided into 5 Local Group galaxies (NGC~598, the Milky Way, IC~10, and the two Magellanic Clouds), and 7 non-Local Group galaxies (the Cartwheel, the Antennae, the Mice, NGC~628, NGC~1058, NGC~5408, and Circinus). The 5 Local Group galaxies have low SFR (the average is $0.36\,{\rm M}_\odot{}\textrm{ yr}^{-1}$, compared to $2.06\,{\rm M}_\odot{}\textrm{ yr}^{-1}$ for the 52 LB05 galaxies) and a total of only one ULX (i.e. an average of 0.2 ULXs/galaxy, compared to 2.29 ULXs/galaxy in the 52 LB05 galaxies; here we are neglecting contamination); whereas the 7 non-local group galaxies have high SFR (average of $5.31\,{\rm M}_\odot{}\textrm{ yr}^{-1}$) and many ULXs (average of 6.86 ULXs/galaxy).

Each of the two non-LB05 sub-groups is clearly different from the sample of LB05 galaxies. However, we still include both of them in our sample, because they allow us to explore regimes that are not well represented by the 52 galaxies in LB05:
 the Local Group galaxies enlarge the number of galaxies with no ULXs, that are likely under-represented because of observation and publication bias; the non-Local Group galaxies provide us with examples (such as the Cartwheel and the Antennae) of relatively rare objects with high SFR.

We note that, even if a galaxy is in the LB05 catalogue, we often use X-ray measurements coming from more recent papers where dedicated analysis have been published: Table~1 includes 27 galaxies for which the raw number of ULXs (N$_{{\rm ULX,\,{}raw}}$, i.e. the observed number of ULXs per galaxy, before the subtraction of contaminating sources)  is taken from LB05, 24 galaxies for which it is taken from LM05, and 13 galaxies for which N$_{{\rm ULX,\,{}raw}}$ is taken from other papers.

\subsubsection{Short description of Table~1}

A full account of the origin of all the data reported in Table~1 will be given in Appendix~B. Here we only provide a basic description of the most important quantities.

The SFRs come from either ultra-violet, H$\alpha{}$, far-infrared, radio measurements,  or an average of these different measurements, depending on the data available for each galaxy.
 Galaxies whose SFR is highly uncertain (a factor of 10 or more), due, e.g., to the presence of an active galactic nucleus (AGN), are not considered in our sample (e.g. NGC~1068, NGC~3690).
Concerning metallicity measurements, their existence is the most restrictive  among our selection criteria, as metallicities are often unavailable. For most galaxies in the sample, we use measurements derived from line intensities in HII regions, translated into abundances with the empirical relation of Pilyugin (2001a; hereafter P01) or Pilyugin \& Thuan (2005; hereafter PT05). If more than one HII region was measured within a galaxy, it is generally possible to derive the metallicity gradient; in such cases, we use the metallicity at the mean distance of ULXs from the centre of their host galaxy ($0.73\,{}R_{25}$, where $R_{25}$ is the radius of the 25th magnitude isophote; see Appendix~B and Liu et al.~2006). When the spectra of HII regions are unavailable, we use X-ray measurements, although they are much less accurate.

The number of expected contaminating sources ($Q$) with apparent luminosities $\ge10^{39}\,{\rm erg\,s}^{-1}$ was calculated on the basis of the Hasinger et~al. (1998) log(N)-log(S) relation, using the distance D and other quantities listed in Appendix~B (e.g. $R_{25}$, which was taken from the RC3 catalogue of de Vaucouleurs et al. 1991).

Finally, the number N$_{{\rm ULX}}$ of ULXs in a galaxy after the subtraction of contaminating sources is simply N$_{{\rm ULX}}={\rm N}_{{\rm ULX, raw}}-Q$.

\begin{figure*}
\center{{
\epsfig{figure=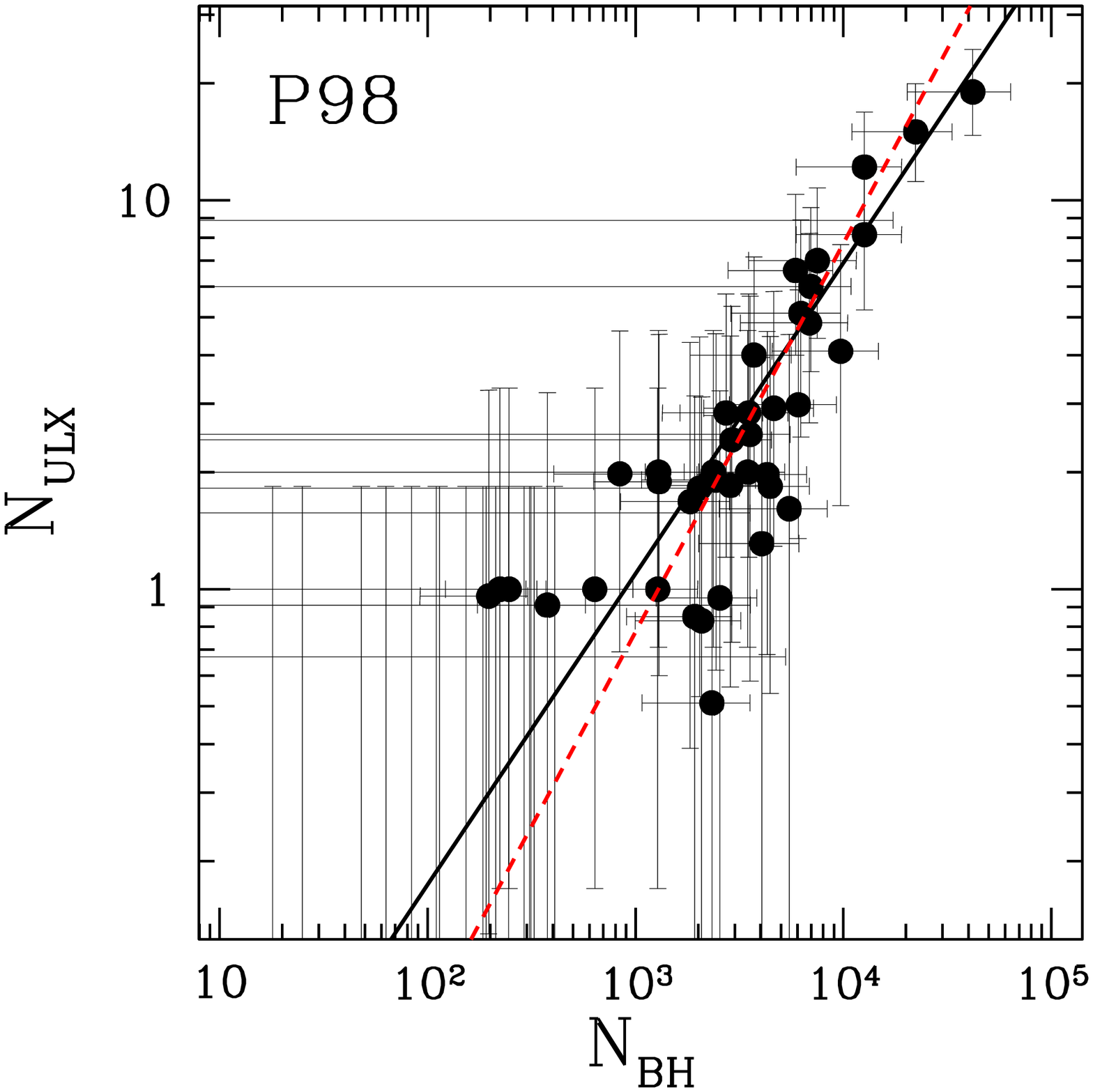,height=7cm} 
\epsfig{figure=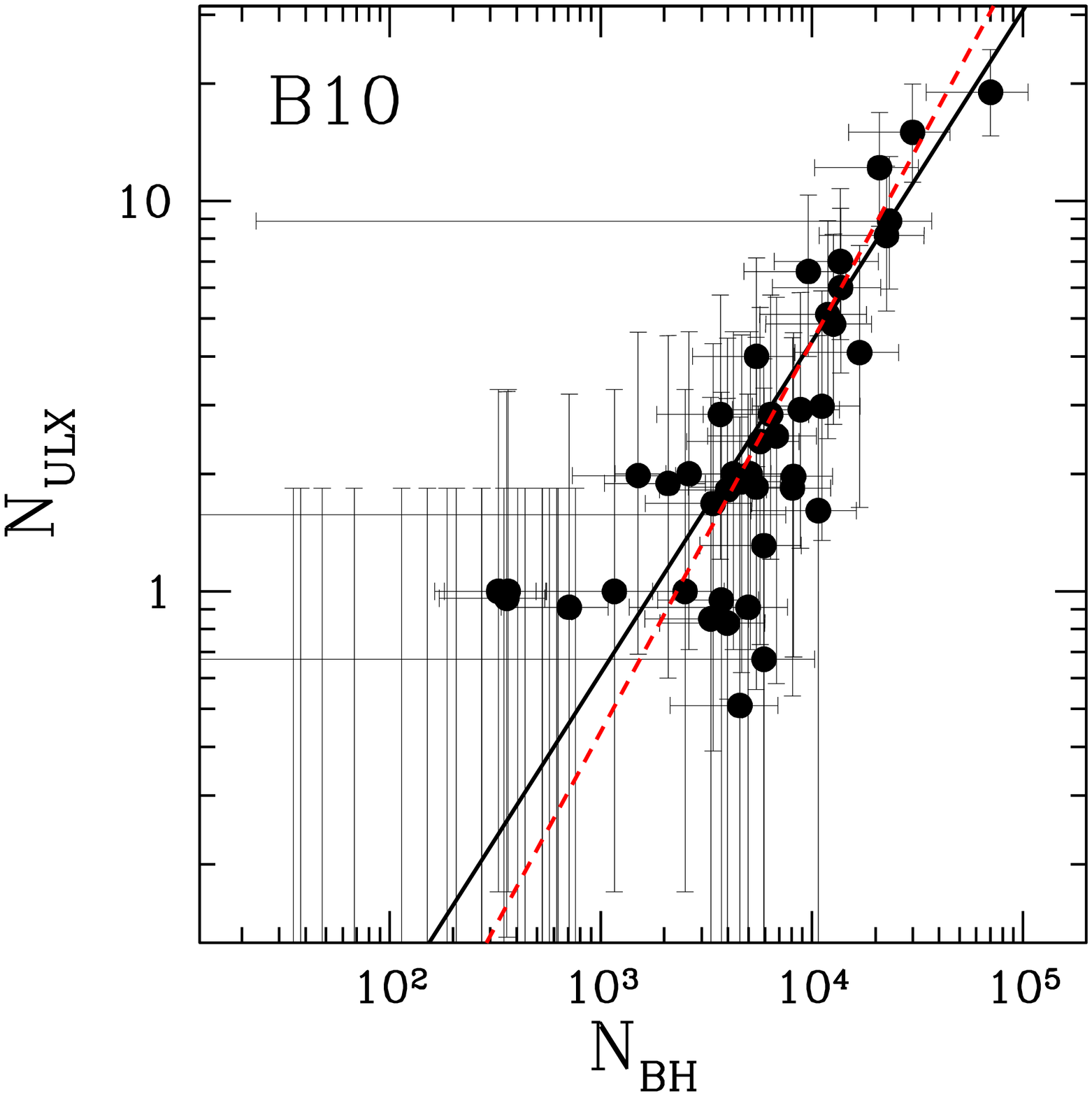,height=7cm} 
}}
\caption{\label{fig:fig2}
 Left-hand panel: Number of observed ULXs per galaxy N$_{\rm ULX}$ versus the number of expected massive BHs per galaxy N$_{\rm BH}$, derived using the models from P98. The adopted IMF is a Kroupa IMF (see Appendix~A for a comparison with the Salpeter IMF). The filled circles are the galaxies listed in Table~1. The solid line is the power-law fit for the entire sample. The dashed line (red on the web) is the power-law fit obtained assuming that the index of the power law is $\beta{}=1$.
The error bars on both the $x-$ and the $y-$ axis are $1-\sigma{}$ errors. The error bars on N$_{\rm BH}$ account for the uncertainty on the SFR and on the metallicity (see Section 2.3 for details).
Right-hand panel: the same as the left-hand panel, but the number of expected massive BHs per galaxy, N$_{\rm BH}$, has been derived using the models from B10.
}
\end{figure*}

\subsection{Uncertainties and fitting procedures}
In the following, we consider various sources of error for the different considered quantities (SFR, $Z$, N$_{\rm ULX}$).

For the SFR, we assume that the relative uncertainty of each value is $\pm 0.5$. This number comes from the analysis of the distribution of the measurements reported in Grimm et al. (2003): the SFR measurements reported in columns 7, 8, 9, and 10 of their table 3 were normalized to the `adopted SFR' reported in column 11 of the same table; the distribution of the resulting values is bell-shaped and centred around 1; furthermore, about 68 per cent of them lies between 0.5 and 1.5. 

We assume that all the metallicity measurements are uncertain by 0.1 dex, which is the typical error associated with metallicity measurements in HII regions (see e.g. P01; Kennicutt et al. 2003; PT05)\footnote{The actual error might be larger for several reasons, such as uncertainties in the metallicity calibrations, or metallicity fluctuations within a single galaxy (even at a given radius).}.

We note that the uncertainty on the SFR affects the error on N$_{\rm BH}$ in a linear way; instead, because of the complicated dependence of N$_{\rm BH}$ upon $Z$ (N$_{\rm BH}$ changes very rapidly with $Z$ for $Z\sim{}0.3-0.4\,{}Z_\odot$, whereas it is almost independent of $Z$ for $Z\gtrsim0.5\,{}Z_\odot$ or $Z\lesssim0.2\,{}Z_\odot$), the uncertainty in $Z$ can result in both very small and very large errors on N$_{\rm BH}$ (this explains the upper limits that can be seen in the left-hand panel of Fig.~\ref{fig:fig2}). 

In the case of ${\rm N}_{{\rm ULX,\,{}raw}}$ we adopt Poisson uncertainties, as they are affected by small-number fluctuations (Grimm et al. 2003). These are estimated through the treatment described in Gehrels (1986).
Instead, in the case of $Q$ we simply assume uncertainties coming from errors upon the $\log({\rm N})-\log({\rm S})$ of Hasinger et~al. (1998).

Finally, the uncertainties on N$_{{\rm ULX}}$ derive from the combination of the two above uncertainties.

The analysis presented in the following sections is based upon correlation coefficients and power-law fits.
Fits were performed by minimizing $\chi^2$. However, since, in most
cases, the quantities on both axes have uncertainties of
similar (relative) magnitude, we keep
both errors into account through the simple procedures summarized in
D'Agostini (2005).

We note that this fact, along with the sizable error bars of the  variables, 
tends to
produce values of $\chi^2$ that are significantly lower than the
number of degrees of freedom (hereafter, dof). 
However, in the low-number regime (that we are
 considering here) the difference between Poissonian
 and Gaussian statistics become significant and hence,  
 whenever possible, we checked our results adopting
 additional fits based on Cash statistics (Cash 1979).
 In such cases,
we further note that i) fits with Cash statistics do {\it not}
 consider the errors on the quantity on the $x$-axis, and ii) in the
cases where the variable on the $y$-axis is N$_{\rm ULX}$, we actually
performed a fit to N$_{\rm ULX,raw}$ with the sum of the contamination
$Q$ to the results of the fitting formula, as required by the Cash
formalism.  The results of the fits based
 on Cash statistics are in reasonably good agreement
 with those of the fits based on $\chi^2$.
 The main difference is that the Cash statistics appears
 to better constrain the parameters and give smaller errors,
 although this follows in part from neglecting uncertainties
 on the $x$-axis. In any case, this result reinforces the
 validity of our analysis.

\section{Results}
\subsection{Expected number of massive BHs versus observed number of ULXs}

\begin{figure}
\center{{
\epsfig{figure=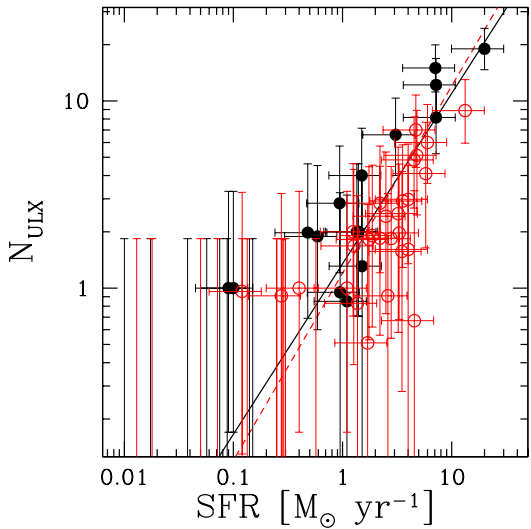,height=7cm} 
}}
\caption{\label{fig:fig3}
N$_{\rm ULX}$ versus the SFR. Filled black circles: galaxies with metallicity $\le0.2\,{}Z_\odot{}$; open circles (red on the web): galaxies with metallicity $>0.2\,{}Z_\odot{}$. Solid line: power-law fit for the entire sample; dashed line (red on the web): power-law fit obtained assuming that the index of the power law is $\delta{}_1=1$. 
 The error bars on both the $x-$ and the $y-$ axis are $1-\sigma{}$ errors (see Section 2.3 for details).}
\end{figure}

\begin{figure}
\center{{
\epsfig{figure=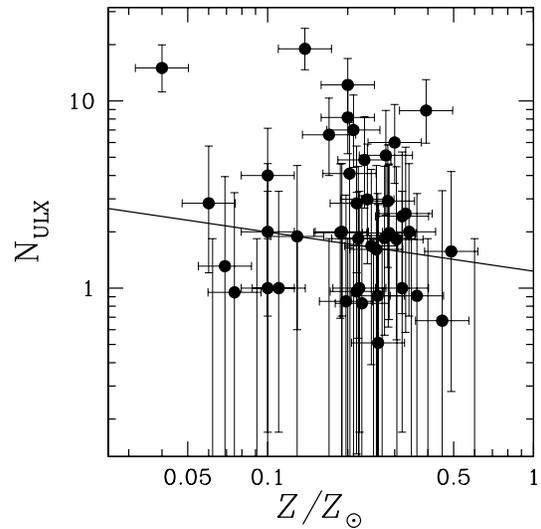,height=7cm} 
}}
\caption{\label{fig:fig4}
N$_{\rm ULX}$ versus $Z$. Filled black circles: entire sample; solid line: power-law fit.  
 The error bars on both the $x-$ and the $y-$ axis are $1-\sigma{}$ errors (see Section 2.3 for details).}
\end{figure}

Table~2 shows the values of N$_{\rm BH}$ and $\epsilon_{\rm BH}$ (the role of $\epsilon_{\rm BH}$ is discussed in Appendix~C) that we derive for the galaxies listed in Table~1,  using the models by P98 and by B10.  In paper~I, we showed that, according to theoretical models (Patruno et al. 2005; Blecha et al. 2006), $\epsilon_{\rm BH}$ should be of the order of $10^{-4}$ for the Kroupa IMF, which is consistent with most values in Table~2 (especially for the models adopting B10 and for galaxies with a large number of ULXs).
In Fig.~\ref{fig:fig2}, we show the observed number of ULXs per galaxy, N$_{\rm ULX}$, as a function of the expected number of massive BHs per galaxy, N$_{\rm BH}$.

From Fig.~\ref{fig:fig2} it is immediately evident that there is a correlation between N$_{\rm BH}$ and N$_{\rm ULX}$, when we adopt the models from both P98 and B10. Such correlation can be expressed as:
\begin{equation}\label{eq:maingoal}
{\rm N}_{\rm ULX}=10^{\gamma{}}\,{}{\rm N}_{\rm BH}^\beta{},
\end{equation}
 where $\beta{}=0.80^{+0.16}_{-0.12}$ ($\beta{}=0.85^{+0.19}_{-0.13}$) and $\gamma{}=-2.36^{+0.45}_{-0.62}$ ($\gamma{}=-2.76^{+0.53}_{-0.76}$), when adopting the models from P98 (B10). Table~3 reports the fits, the results of the $\chi{}^2$ analysis, the Pearson correlation coefficients (r), and the probability p$_{\rm r}$ to obtain\footnote{Such probability is based on the assumption that the two variables whose correlation is being tested are normally distributed. This is likely not the case for our data; however, the value of p$_{\rm r}$ can still be retained as an indicator of the strength (or weakness) of a correlation.}
a correlation coefficient with an absolute value larger than $|{\rm r}|$ if the two variables were uncorrelated.

The correlation coefficient r is very high, and the probability of finding a value larger than $|{\rm r}|$, if the two variables were uncorrelated and normally distributed, is almost zero.
The values of the $\chi{}^2$ associated with the best fits of the ${\rm N}_{\rm BH}-{\rm N}_{\rm ULX}$ relation are $\chi{}^2\lesssim{}12$, with 62 degrees of freedom (dof). The $\chi{}^2$ is almost as low when an index $\beta{}=1.00$ is assumed.  According to the F$-$test (see paragraph 10.2 in Bevington 1969), the above fit with two parameters (see lines~1 and 3 of Table~3) is better than a fit with fixed index $\beta{}=1$ (see lines~2 and 4 of Table~3) only 
at 91 and 80 per cent confidence level 
in the case of P98 and B10, respectively. This implies that, especially for B10, the fit with two parameters is not a significant improvement.
This suggests that the ${\rm N}_{\rm BH}-{\rm N}_{\rm ULX}$ correlation is a linear relation, although deviations from a linear behaviour are possible\footnote{The scaling between N$_{\rm BH}$ and N$_{\rm ULX}$ might be non-linear because N$_{\rm ULX}$ may depend also on properties that do not affect N$_{\rm BH}$, such as the probability of finding sufficiently massive companion stars. Thus, a better understanding of the properties of HMXBs that can power ULXs is needed, in order to refine our model.}.

Furthermore, the slopes are very similar for the two considered models of stellar evolution. The main difference between them is the normalization, as the values of N$_{\rm BH}$ obtained assuming P98 are generally a factor of $1.5-2$ lower than the values obtained using B10.

 Finally, from Tables~$2-3$ and from Fig.~\ref{fig:fig2}, it appears that the galaxies in our sample that do not host any ULXs have very low values of N$_{\rm BH}$: according to our model, they are consistent with having one or zero ULXs.

In conclusion, Fig.~\ref{fig:fig2} and Tables~2 and 3 indicate that our model predicts a number of massive BHs that correlates with the number of ULXs per galaxy, independently of the adopted stellar evolution scenario. Such correlation  supports the hypothesis that (all or a large fraction of) the ULXs are connected with massive ($m_{\rm BH}\ge{}25\,{}{\rm M}_\odot{}$) BHs.

\begin{table*}
\begin{center}
\caption{Parameters of the  power-law fits and $\chi{}^2$.} \leavevmode
\begin{tabular}[!h]{llllllll}
\hline
$x-$ axis
& $y-$ axis
& Model
& Sample $^{\rm a}$
& Index $^{\rm b, c}$
& Normalization $^{\rm c}$
& $\chi{}^2/{\rm dof}$ $^{\rm d}$
& r (p$_{\rm r}$) $^{\rm e}$\\
\hline
${\rm N}_{\rm BH}$  & ${\rm N}_{\rm ULX}$ & P98 & all & $0.80^{+0.16}_{-0.12}(0.86\pm0.07)$ &   $-2.36^{+0.45}_{-0.62}(-2.64\pm0.28)$ &  $8.7/62$   & 0.90 (2$\times{}10^{-23}$)\\ 
${\rm N}_{\rm BH}$  & ${\rm N}_{\rm ULX}$ & P98 & all & $1.00(1.00)$          &   $-3.11\pm{}0.07(-3.17\pm0.04)$ & $10.0/63$   & 0.90 (2$\times{}10^{-23}$)\\   
${\rm N}_{\rm BH}$  & ${\rm N}_{\rm ULX}$ & B10 & all & $0.85^{+0.19}_{-0.13}(0.90\pm0.07)$ &   $-2.76^{+0.53}_{-0.76}(-3.00\pm0.30)$ & $11.1/62$   & 0.93 (1$\times{}10^{-27}$)\\   
${\rm N}_{\rm BH}$  & ${\rm N}_{\rm ULX}$ & B10 & all & $1.00(1.00)$          &   $-3.36\pm{0.07}(-3.41\pm0.04)$ & $11.8/63$   & 0.93 (1$\times{}10^{-27}$)\vspace{0.2cm}\\   

SFR   & ${\rm N}_{\rm ULX}$         & --  & all &  $0.91^{+0.25}_{-0.15}(0.90\pm0.08)$ &  $0.13^{+0.10}_{-0.14}(0.09\pm0.06)$  & $17.7/62$   & 0.88 (4$\times{}10^{-22}$)\\
SFR   & ${\rm N}_{\rm ULX}$         & --  & all &  $1.00 (1.00)$                 &  $0.08\pm{}0.06(0.02\pm0.04)$         &   $17.8/63$ & 0.88 (4$\times{}10^{-22}$) \\
SFR   & ${\rm N}_{\rm ULX}$           & --  & lowZ &  $0.75^{+0.20}_{-0.13}(0.80\pm0.08)$  &  $0.39^{+0.09}_{-0.13}(0.32\pm0.07)$  &  $4.4/22$ & 0.93 (8$\times{}10^{-11}$)\\ 
SFR   & ${\rm N}_{\rm ULX}$           & --  & highZ &  $0.83^{+0.37}_{-0.22}(0.94\pm0.15)$  &  $0.05^{+0.13}_{-0.20}(-0.06\pm0.10)$  &   $6.4/38$ & 0.88 (6$\times{}10^{-14}$) \vspace{0.2cm}\\

$Z/Z_\odot{}$  & ${\rm N}_{\rm ULX}$ & -- & all & $-0.21\pm{}0.27(-0.55\pm0.15)$ &  $0.09\pm{}0.20(-0.04\pm0.12)$  &    $86.0/62$ & $-0.16$ (2$\times{}10^{-1}$)\\

$Z/Z_\odot{}$  & ${\rm N}_{\rm ULX}$ & -- & all & $0.00(0.00)$ &  $0.23\pm{}0.05(0.36\pm0.04)$  &    $86.6/63$ & $-0.16$ (2$\times{}10^{-1}$)\\

$Z/Z_\odot{}$  & ${\rm N}_{\rm ULX}/{\rm SFR}$ & -- & all & $-0.55\pm{}0.23$ &  $-0.37\pm{}0.18$  &    $10.4/62$  & $-0.30$ (2$\times{}10^{-2}$)\\

$Z/Z_\odot{}$  & ${\rm N}_{\rm ULX}/{\rm SFR}$ & -- & all & $0.00$ &  $-0.03\pm{}0.07$  &    $14.7/63$  & $-0.30$ (2$\times{}10^{-2}$)\vspace{0.2cm}\\

SFR  & ${\rm N}_{\rm BH}$$^{\rm f}$ & P98 & all & $0.96\pm{}0.06$ &   $3.19\pm{}0.04$  &  $13.8/62$ & 0.82 (7$\times{}10^{-17}$)\\
SFR  & ${\rm N}_{\rm BH}$$^{\rm f}$ & B10 & all & $0.97\pm{}0.05$ &   $3.44\pm{}0.04$  &  $6.3/62$  & 0.95 (6$\times{}10^{-33}$) \vspace{0.2cm}\\

$Z/Z_\odot{}$ & ${\rm N}_{\rm BH}$$^{\rm f}$ & P98 & all & $-0.19\pm{}0.29$ &  $1.41^{+0.23}_{-0.26}$  &  $153.9/62$ & $-0.23$ (7$\times{}10^{-2}$) \\
$Z/Z_\odot{}$ & ${\rm N}_{\rm BH}$$^{\rm f}$ & B10 & all & $0.05^{+0.30}_{-0.27}$  &  $1.85\pm{}0.24$  &  $183.2/62$ & $-0.11$ (4$\times{}10^{-1}$) \vspace{0.2cm}\\

$Z/Z_\odot{}$ & ${\rm N}_{\rm BH}/{\rm SFR}$ $^{\rm f}$ & P98 & all & $-0.60\pm{}0.07$ &  $2.79\pm{}0.05$ &   $9.4/62$ & $-0.96$ (2$\times{}10^{-37}$)\\

$Z/Z_\odot{}$ & ${\rm N}_{\rm BH}/{\rm SFR}$ $^{\rm f}$ & B10 & all & $-0.34^{+0.02}_{-0.05}$ & $3.22\pm{}0.04$  & $17.0/62$ & $-0.98$ (3$\times{}10^{-48}$)\vspace{0.2cm}\\

\noalign{\vspace{0.1cm}}
\hline
\end{tabular}
\end{center}
\footnotesize{
The SFRs used by the fitting procedure are in units of M$_\odot{}$ yr$^{-1}$.
$^{\rm a}$ The sample adopted for the fits is referred to as `all' when all the galaxies in Table~1  are considered, and as lowZ (highZ) when only the galaxies with $Z\le{}0.2\,{}Z_\odot{}$ ($Z>0.2\,{}Z_\odot{}$) are considered.
$^{\rm b}$  When the index is equal to 1.00 or to 0.00, without error, it means that it has been fixed to such value.
$^{\rm c}$ Values in parenthesis are the fitting parameters obtained with Cash statistics (where applicable), rather than $\chi^2$; such results do not take into account the uncertainty on the quantity on the $x$-axis.
$^{\rm d}$ $\chi{}^2/{\rm dof}$ is the $\chi{}^2$ divided by the degrees of freedom (dof).
$^{\rm e}$  r is the Pearson correlation coefficient, and p$_{\rm r}$ is the probability of finding a value larger than $|{\rm r}|$, if the two variables were uncorrelated and normally distributed. $^{\rm f}$~The fits listed in rows 13--18 refer entirely to the theoretical model. We report them only for a comparison between the statistical fluctuations of the theoretical calculations and those of the observational data.}
\end{table*}

\subsection{Dependence on the SFR and on the metallicity}
Our model is based on two key parameters: SFR and metallicity. In this Subsection we discuss their role in more detail.

\subsubsection{Star formation rate}
As mentioned in the Introduction, the existence of a correlation between SFR and number of bright X-ray sources per galaxy is well established (Grimm et al. 2003; Ranalli et al. 2003; Gilfanov et al. 2004a,b,c).  From Fig.~\ref{fig:fig3}, it is evident that our sample of galaxies follows the same correlation, which can be  fit as:
\begin{equation}\label{eq:ULXSFR}
{\rm N}_{\rm ULX}=10^{\zeta{}}\,{}\left(\frac{{\rm SFR}}{{\rm M}_\odot{}\textrm{ yr}^{-1}}\right)^{\delta{}},
\end{equation}
  where $\delta{}=0.91^{+0.25}_{-0.15}$ and $\zeta{}=0.13^{+0.10}_{-0.14}$ (with $\chi{}^2\sim{}18$ for 62 dof, see Table~3), if we consider all the galaxies in the sample.  We note that the index of the correlation ($\delta{}$) is smaller than 1, but consistent with 1. Furthermore, the value of the $\chi{}^2$ is very low ($\chi{}^2\sim{}18$ with 63 dof), even when we assume $\delta{}=1.00$.  According to the F$-$test, the above fit with two parameters (see line~5 of Table~3) is better than a fit with fixed index $\delta{}=1$ (see line~6  of Table~3) 
only at 48 per cent confidence level. This means that the fit with two parameters is not an improvement.

 If we impose  $\delta{}=1.00$, we obtain N$_{\rm ULX}\simeq{}1.20^{+0.18}_{-0.15} \textrm{ SFR}$. 
Likewise, Grimm et al. (2003) find an almost linear relation between the SFR and the number of X-ray sources with $L_{\rm X}\ge{}2\times{}10^{38}$ erg s$^{-1}$. Our best-fitting slope in the linear relation ($1.20^{+0.18}_{-0.15}$) is a factor of $2.4\pm0.4$ lower than the value ($2.9\pm0.23$) reported by Grimm et al. (2003); given the different luminosity ranges ($L_{\rm X}\ge{}2\times{}10^{38}$ erg s$^{-1}$ and $L_{\rm X}\ge{}10^{39}$ erg s$^{-1}$ in Grimm et al. 2003 and in this Paper, respectively), this is expected. In fact, equation~(7) of Grimm et~al. (2003) predicts a factor $3.0\pm0.5$ difference between the normalizations of the correlations in the two luminosity ranges, consistent with our result (the uncertainty is due to the error on the slope of the luminosity function given in Equation~6 of Grimm et~al. 2003).

We remind also that, in our model, we imposed that ${\rm N}_{\rm BH}$ scales linearly with the SFR (see equation~\ref{eq:norm}), in agreement with the observational behaviour of N$_{\rm ULX}$. Lines 13$-$14 of Table~3 indicate that our fitting procedure infers a slope which is consistent with the theoretical value within a $\pm{}0.06$ uncertainty. The normalization for the model by P98 is a factor of $\sim{}1.5-2$ lower than that of the model by B10.

\subsubsection{Metallicity}
In Paper~I we proposed that there may be an anticorrelation between the observed number of ULXs (N$_{\rm ULX}$) and the metallicity ($Z$), but our sample was too small to confirm such anticorrelation. Our current sample is more than three times larger than the one in Paper~I.

In the current paper, a first argument in favour of the role played by the metallicity comes from Fig.~\ref{fig:fig3} (i.e. N$_{\rm ULX}$ versus SFR): it is evident that the galaxies with $Z>0.2\,{}Z_\odot{}$ (open circles) lie mostly below the global fit, whereas the galaxies with $Z\le{}0.2\,{}Z_\odot{}$ (filled circles) lie preferentially above the global fit.
This means that a metal-poor ($Z\le{}0.2\,{}Z_\odot{}$) galaxy tends to have more ULXs with respect to a relatively metal-rich ($Z>0.2\,{}Z_\odot{}$) galaxy with the same SFR. In quantitative terms, Table~3 shows that, if we split our sample according to metallicity and fit the SFR$-$N$_{\rm ULX}$ relation, we obtain similar slopes, but quite different values of the  normalization for the high-$Z$ sample ($\zeta_1 = 0.05^{+0.13}_{-0.20}$) and for  the low-$Z$ sample ($\zeta_1 = 0.39^{+0.09}_{-0.13}$). The difference ($0.34^{+0.22}_{-0.18}$; or $0.32\pm0.14$ if the slope is fixed to $\delta=1.00$) is not consistent with 0 at a significance level slightly below $2\,{}\sigma$, while it is roughly consistent with the ratio of the expected numbers of massive BHs below and above $Z\sim{}0.2\,{}Z_{\odot{}}$ (see Fig.~\ref{fig:fig1}).
This might be a further indication that the SFR is not the only ingredient of the correlation between ${\rm N}_{\rm BH}$ and ${\rm N}_{\rm ULX}$ and that the second ingredient might actually be the metallicity, as proposed in our model.

However, we cannot conclude from   Fig.~\ref{fig:fig4} and from Table~3 that such anticorrelation exists. There is a very weak trend that can be expressed as
\begin{equation}\label{eq:ULXZ}
{\rm N}_{\rm ULX}=10^{\theta{}}\,{}(Z/Z_\odot{})^{\eta{}},
\end{equation}
where $\eta{}=-0.21\pm{}0.27$ and $\theta{}=0.09\pm{}0.20$. Such trend is not statistically significant, 
as $\chi{}^2=86$ with 62 dof 
  and the correlation coefficient is r$\lesssim{}-0.2$, with p$_{\rm r}=0.2$ (Table~3).

\begin{figure*}
\center{{
\epsfig{figure=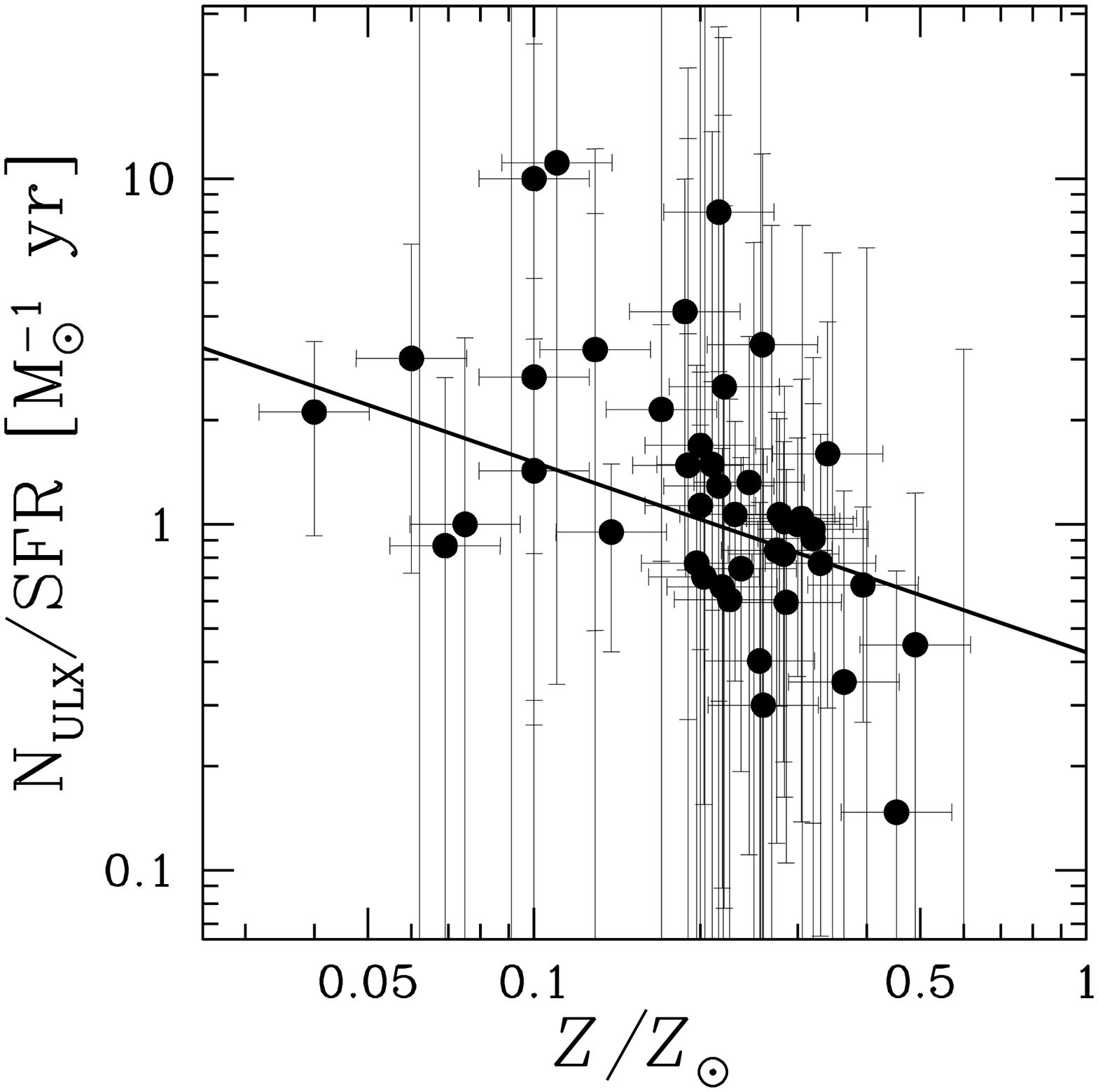,height=5.5cm} 
\epsfig{figure=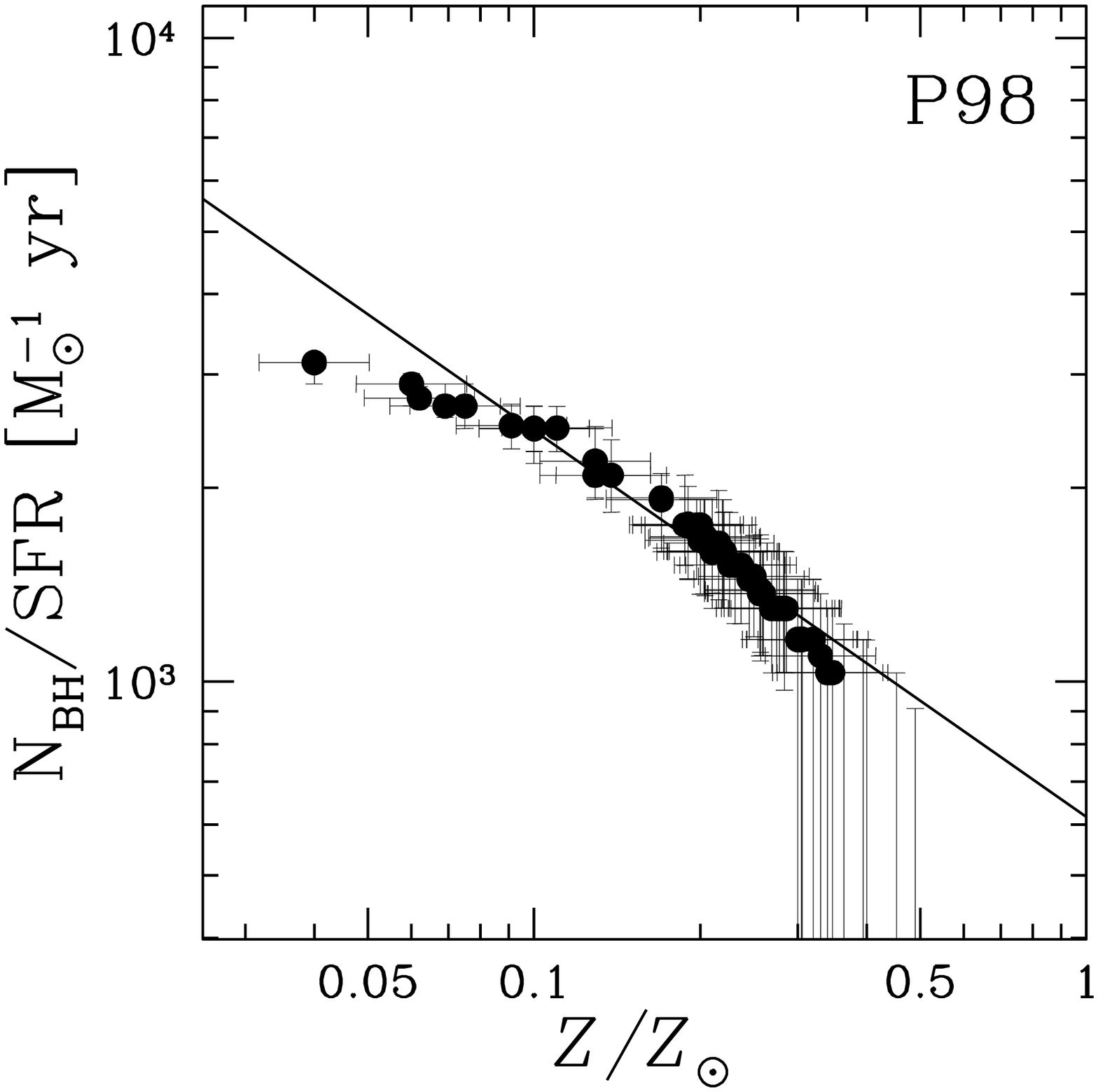,height=5.5cm}  
\epsfig{figure=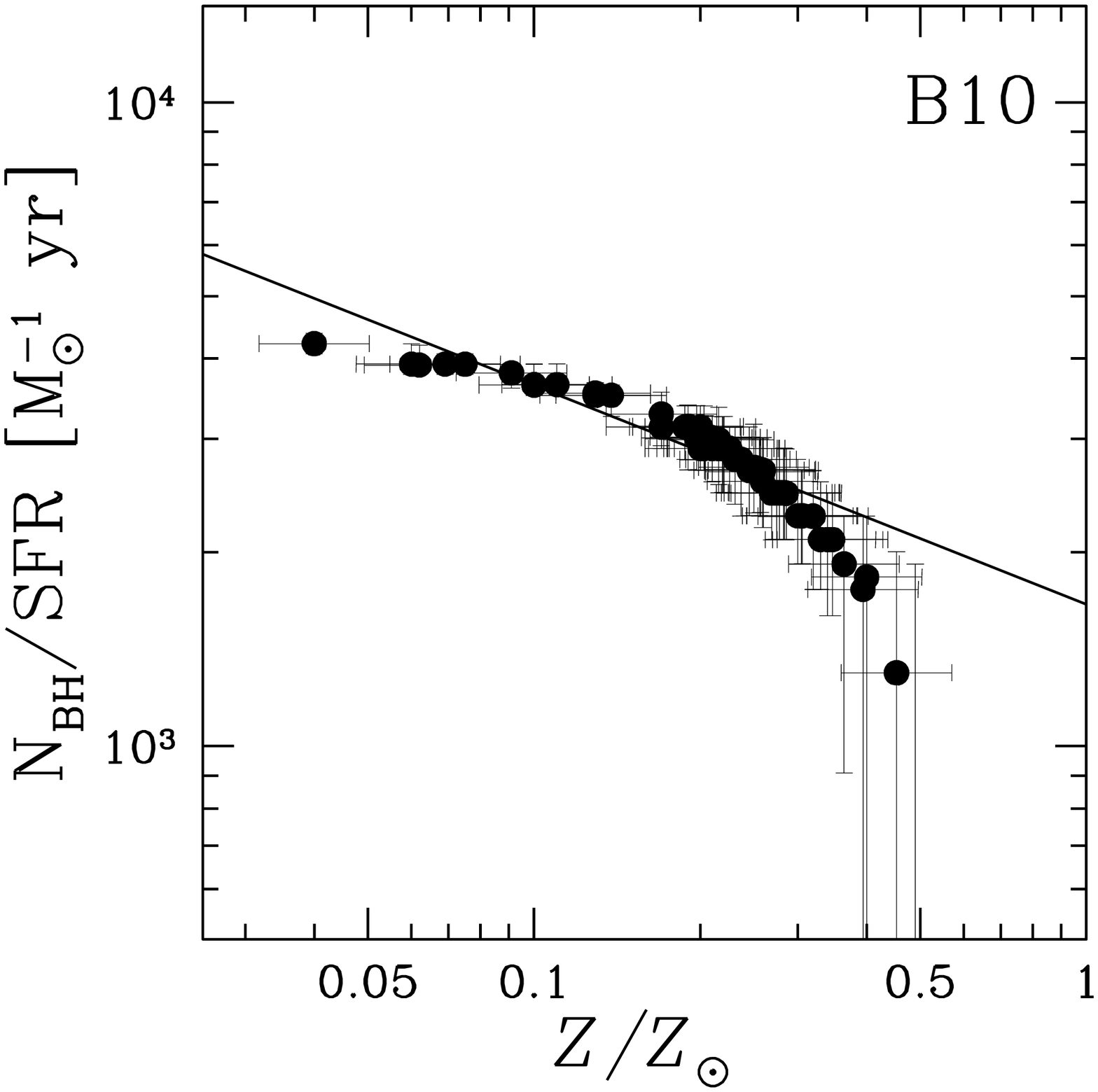,height=5.5cm} 
}}
\caption{\label{fig:fig5}
 From left to right: N$_{\rm ULX}/{\rm SFR}$ versus $Z$,  N$_{\rm BH}/{\rm SFR}$ versus $Z$ for P98, N$_{\rm BH}/{\rm SFR}$ versus $Z$ for B10.
 Filled black circles: entire sample. 
Solid lines: power-law fit.  
 The error bars on both the $x-$ and the $y-$ axis are $1-\sigma{}$ errors. Central and right-hand panel: the error bars on N$_{\rm BH}/{\rm SFR}$ account for the uncertainty on the metallicity (see Section 2.3 for details).}
\end{figure*}

Similarly to equation~(\ref{eq:ULXZ}), there is no significant anticorrelation, in our model, between N$_{\rm BH}$ and $Z$ (lines 15-16 of Table~3). In particular, the values of the $\chi{}^2$ for the best fits are 154 and 183 for P98 and B10, respectively, with 62 dof (see lines 15 and 16 of Table~3). The correlation coefficients are low: r$=-0.2$ and $=-0.1$ for P98 and B10, respectively. This happens because, in our model, the dependence of N$_{\rm BH}$ on the SFR dominates with respect to the dependence of N$_{\rm BH}$ on the metallicity.

To better highlight the possible effects of metallicity, we plot the number of ULXs per galaxy N$_{\rm ULX}$ normalized to the SFR versus the metallicity (left-hand panel of Fig.~\ref{fig:fig5}). The fit of the left-hand  panel in Fig.~\ref{fig:fig5} gives the following results.
\begin{equation}\label{eq:ULXSFRZ}
\left({\rm N}_{\rm ULX}\,{}\,{}\frac{{\rm M}_\odot{}\textrm{ yr}^{-1}}{\rm SFR}\right)=10^{\kappa{}_1}\,{}(Z/Z_\odot{})^{\iota{}_1},
\end{equation}
 where $\iota{}_1=-0.55\pm{}0.23$ and $\kappa{}_1=-0.37\pm{}0.18$, with $\chi{}^2\sim{}10$ for 62 dof (and r$=-0.3$).
Thus, there is marginal evidence of an anticorrelation between $Z$ and N$_{\rm ULX}/{\rm SFR}$.  
 According to the F$-$test, the above fit with two parameters (see line~11 of Table~3) is significantly better 
(at a 96 per cent significance level) than a fit with fixed index $\iota{}_1=0$ (i.e. the case of no correlation) and with only one free parameter ($\kappa{}_1=-0.03\pm{}0.07$, line~12 of Table~3). 
This result supports the hypothesis of an anti-correlation between N$_{\rm ULX}/{\rm SFR}$ and $Z$.

As in the case of the comparison with N$_{\rm ULX}$, the role of metallicity in our model can be better appreciated if we remove the contribution from the SFR. In fact, the plot of N$_{\rm BH}/{\rm SFR}$ versus $Z$ (central and right-hand panels of Fig.~\ref{fig:fig5}) shows quite clearly the effect of metallicity in equation~(\ref{eq:totnum}).  We know from equation~(\ref{eq:totnum}) that the relation between ${\rm N}_{\rm BH}/{\rm SFR}$ and $Z$ is not a power-law. However, in order to compare it with equation~(\ref{eq:ULXSFRZ}), we can approximate it as:
\begin{equation}\label{eq:BHSFRZ}
\left({\rm N}_{\rm BH}\,{}\,{}\frac{{\rm M}_\odot{}\textrm{ yr}^{-1}}{\rm SFR}\right)=10^{\kappa{}_2}\,{}(Z/Z_\odot{})^{\iota{}_2},
\end{equation}
where $\iota{}_2=-0.60\pm{}0.07$ ($-0.34^{+0.02}_{-0.05}$) and $\kappa{}_2=2.79\pm{}0.05$ ($3.22\pm{}0.04$) for P98 (B10). The statistical significance of such correlations is given in lines 17$-$18 of Table~3. $\iota{}_1$ is consistent with $\iota{}_2$ in the case of P98 and marginally consistent with $\iota{}_2$ in the case of B10. Thus, we can conclude that the anticorrelation in the central and right-hand panels of Fig.~\ref{fig:fig5} is consistent with that in the left-hand panel of the same Figure, although the error bars are quite large.

The fact that the anticorrelation between $Z$ and N$_{\rm ULX}$  only emerges after removing the effect of the SFR (i.e. considering N$_{\rm ULX}/$SFR versus $Z$) clearly shows that the SFR (and not the metallicity) is the dominant factor in determining the number of ULXs in a given galaxy\footnote{The importance of the SFR is somewhat amplified by the larger spread in the SFR values with respect to the metallicity ones.}.
However, the metallicity plays a crucial role in enabling the formation of ULXs, as massive BHs can be produced only in star-forming regions where $Z$ is sufficiently low.

We emphasize that, in our model, the effect of the metallicity on the formation of massive BHs is mainly a threshold effect: in the models by B10 (see their fig.~1 and their equation 11) massive ($m_{\rm BH}\ge{}25\,{}{\rm M}_\odot{}$)  BHs form only for $Z\le{}0.47\,{}Z_\odot{}$. If the metallicity $Z$ is below this threshold, there is quite a small  spread in  $m_{\rm prog}$ (defined in equation~\ref{eq:totnum}) and thus in N$_{\rm BH}$ for various metallicities (at fixed SFR): adopting the model by B10, $m_{\rm prog}$ ranges from $68\,{}{\rm M}_\odot{}$ for $Z=0.04\,{}Z_\odot{}$ (the Antennae) to $96\,{}{\rm M}_\odot{}$ for $Z=0.45\,{}Z_\odot{}$ (NGC~2442).
Furthermore, as N$_{\rm BH}$ scales linearly with the SFR, the effect of metallicity on N$_{\rm BH}$ is  small with respect to the effect of the SFR, for $Z<0.47\,{}Z_\odot{}$.

Instead, if $Z>0.47\,{}Z_\odot{}$, massive BHs cannot form from the direct collapse of massive stars. This might be a problem for our model, as ULXs exist also in  a few galaxies with $Z>0.47\,{}Z_\odot{}$:  NGC~4501 (see Table~1) is the only galaxy, for which we found SFR, $Z$ and X-ray data, that has N$_{{\rm ULX,\,{}raw}}>0$ and $Z>0.47\,{}Z_\odot{}$. In such a case, the contamination from foreground/background sources ($Q=0.43^{+0.11}_{-0.09}$) is not completely negligible; nonetheless it is insufficient to explain the 2 observed ULXs.
However, there are several reasons for which this might happen.

  First, as the metallicity is uncertain by $\sim{}0.1$ dex,  values of $Z$ below $0.47\,{}Z_\odot{}$ are well within $1\,{}\sigma{}$ error ranges for NGC~4501 ($1-\sigma$ lower limit on $Z$ is
$0.39\,{}Z_\odot{}$ for NGC~4501). Thus, the upper limits of N$_{\rm BH}$ for NGC~4501 are larger than zero.

Second, in the case of large spiral galaxies (such as NGC~4501) our choice of taking the metallicity at $R=0.73\,{}R_{25}$, although justified, is rather simplistic: in NGC~4501 the metallicity gradient is such that $Z$ goes below $0.4\,{}Z_\odot$ at $R\le{} R_{25}$, so that there exists a small region where massive BHs might form, even in the P98 model. Thus, uncertainties in the metallicity determination and spatial abundance fluctuations might further help in this respect.

Third, in this Paper we considered only BHs with $m_{\rm BH}\ge25\,{}{\rm M}_\odot$ as engines of the ULXs, but slightly less massive BHs might still power many low-luminosity ULXs, provided that there is a certain level of beaming or super-Eddington accretion (e.g. King 2008): for example, if the metallicity is $Z\lesssim{}0.57\,{}Z_\odot{}$ BHs with m$_{\rm BH}\sim{}20\,{}{\rm M}_{\odot{}}$ can still form (B10).
 It is worth mentioning that
  NGC~4501 hosts relatively faint ULXs, that can be explained without invoking massive BHs, nor strong beaming/super-Eddington accretion.

Finally,
the metallicity needed in our model is that of the molecular clouds before the pollution from the first supernovae associated with the parent cluster, as massive stars ($>40\,{}{\rm M}_\odot{}$) collapse into BHs before the explosion of such supernovae. Thus, the metallicity measured today is likely higher than the value we should consider in our model, especially for post-starburst galaxies.

\section{Conclusions}
Low-metallicity ($Z\lesssim{}0.4\,{}Z_\odot{}$) massive ($\gtrsim{}40\,{}{\rm M}_\odot{}$) stars are expected to produce massive remnants ($25\le{}m_{\rm BH}/{\rm M}_\odot{}\le{}80$, H03, B10) at the end of their evolution. Such massive BHs might power a large fraction of the observed ULXs in low-metallicity galaxies. In this Paper, we derived the number of massive BHs (N$_{\rm BH}$) that are expected to form in a galaxy, via this mechanism, in the same time period that they could have a massive ($\gtrsim{}15$ M$_\odot{}$) donor companion. We find that N$_{\rm BH}$ correlates  well with the observed number of ULXs per galaxy (N$_{\rm ULX}$). The slope of such correlation does not depend significantly either on the IMF or on the adopted stellar evolution model. The IMF and the stellar-evolution models affect only the normalization of  N$_{\rm BH}$ (the spread is generally less than a factor of 2).
 We stress that the stellar-evolution models adopted in this Paper neglect some important effects, such as the rotation and the possible influence of binary evolution. The final mass of the remnant is likely affected by the fact that the massive progenitor of the BH is fast rotating or that it resides in a binary, where additional mass loss is possible (see e.g. H03).
Accounting for the probability that the progenitor of the massive BH is in a binary system likely introduces additional uncertainties.

 In addition, the model considered here does not include the possibility of pair instability supernovae (PISNs). PISNs are predicted to occur at very low metallicity, in the case of very massive stars ($\ge{}140$~M$_\odot${}), and  lead to the complete disruption of the star (HW02; H03). 
Pulsational pair instability may occur also for smaller stellar masses ($100-140$~M$_\odot{}$) and/or for larger metallicity, but it does not lead to a PISN, and a massive BH can form via direct collapse. 
Thus, PISNs probably do not play a role for stars with metallicity $Z\gtrsim{}0.01\,{}Z_\odot$ and mass $<140$~M$_\odot{}$ (H03). On the other hand, even assuming (as a strong upper limit) that all stars with mass $\ge{}100\,{}M_\odot{}$ do not leave any remnant, due to a PISN, our estimates of N$_{\rm BH}$ change by less than 10 per cent.

The model described in this Paper is 
consistent with the observed correlation between the number of ULXs and the SFR, as well as with the fact that ULXs are found preferentially in low-metallicity environments.
Furthermore, this model is a natural extension of the scenario described by Grimm et al. (2003). In fact, Grimm et al. (2003) find a correlation between the SFR and the number of X-ray sources (not necessarily ULXs) per galaxy, and they explain it with the correlation between the SFR and the number of HMXBs powered by stellar BHs. In this Paper we suggest that this correlation still holds for the ULXs, as the ULXs (or most of them) can be powered by massive BHs formed by the collapse of  massive metal-poor  stars.

Our model predicts the existence of a dependence of N$_{\rm ULX}$ on $Z$ (and the data  suggest it, too), when the dominant effect due to the SFR is removed. 
Unfortunately, the statistical uncertainty of such dependence  is still quite high, due to the dearth and to the inhomogeneity of the data.

In particular, there is a large inhomogeneity in the metallicity measurements. Moreover, the metallicity needed in our model is that of the molecular clouds before the pollution from the first supernovae associated with the parent clusters, as very massive stars collapse into BHs before such supernovae. Thus, the metallicity measured today is likely higher than the value we should consider in our model. Only for some types of galaxies, such as the ring galaxies, where the star-formation history has a clear connection with the geometry of the system, is it possible to measure a pre-starburst value of $Z$, suitable for our purposes. A possible way to reduce this problem and to check our model is to take new measurements of the local metallicity in the neighbourhoods of the observed ULXs, or even in the nebula associated with the ULXs (Ripamonti et al., in preparation).

\section*{Acknowledgments}
We thank A.~Wolter, G.~Trinchieri, F.~Pizzolato, R.~Decarli,
G. Ghirlanda, and P.~Marigo for useful discussions.  We thank the
referee for his/her critical reading of the manuscript. MM acknowledges
support from the Swiss National Science Foundation, project number
200020-109581/1 and from the Forschungskredit Fellowship 2008 of the
University of Z\"urich.  MC and LZ acknowledge financial support from
INAF through grant PRIN-2007-26. AB acknowledges ASI INAF contract
ASI-INAF I/016/07/0.  This research has made use of NASA's
Astrophysics Data System Bibliographic Services, of the NASA/IPAC
Extragalactic Database (NED; which is operated by the Jet Propulsion
Laboratory, California Institute of Technology, under contract with
the National Aeronautics and Space Administration), and of data
obtained from the High Energy Astrophysics Science Archive Research
Center (HEASARC; which is provided by NASA's Goddard Space Flight
Center).

\appendix
\section{Comparison between Kroupa and Salpeter IMF}
\begin{figure*}
\center{{
\epsfig{figure=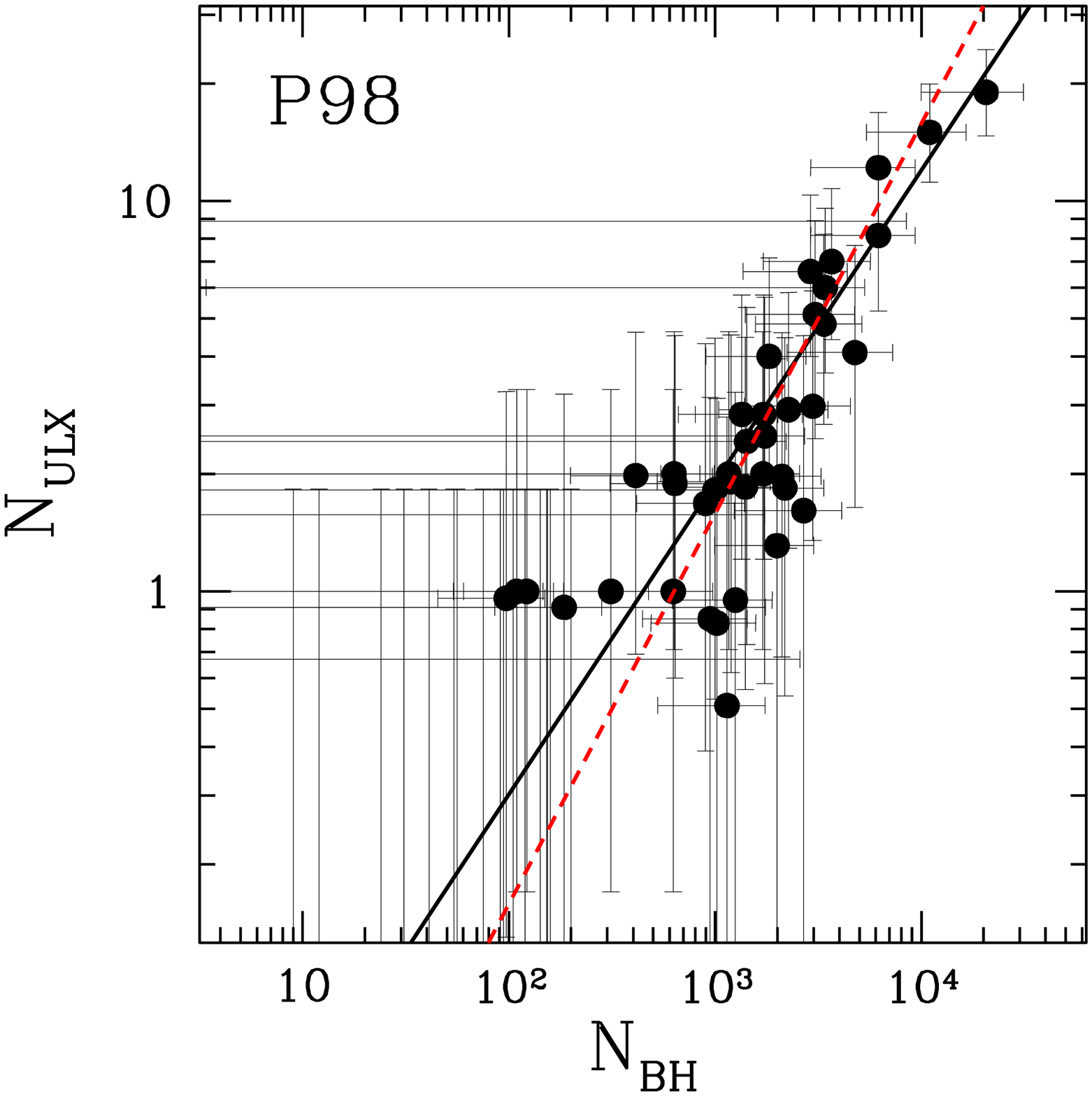,height=7cm} 
\epsfig{figure=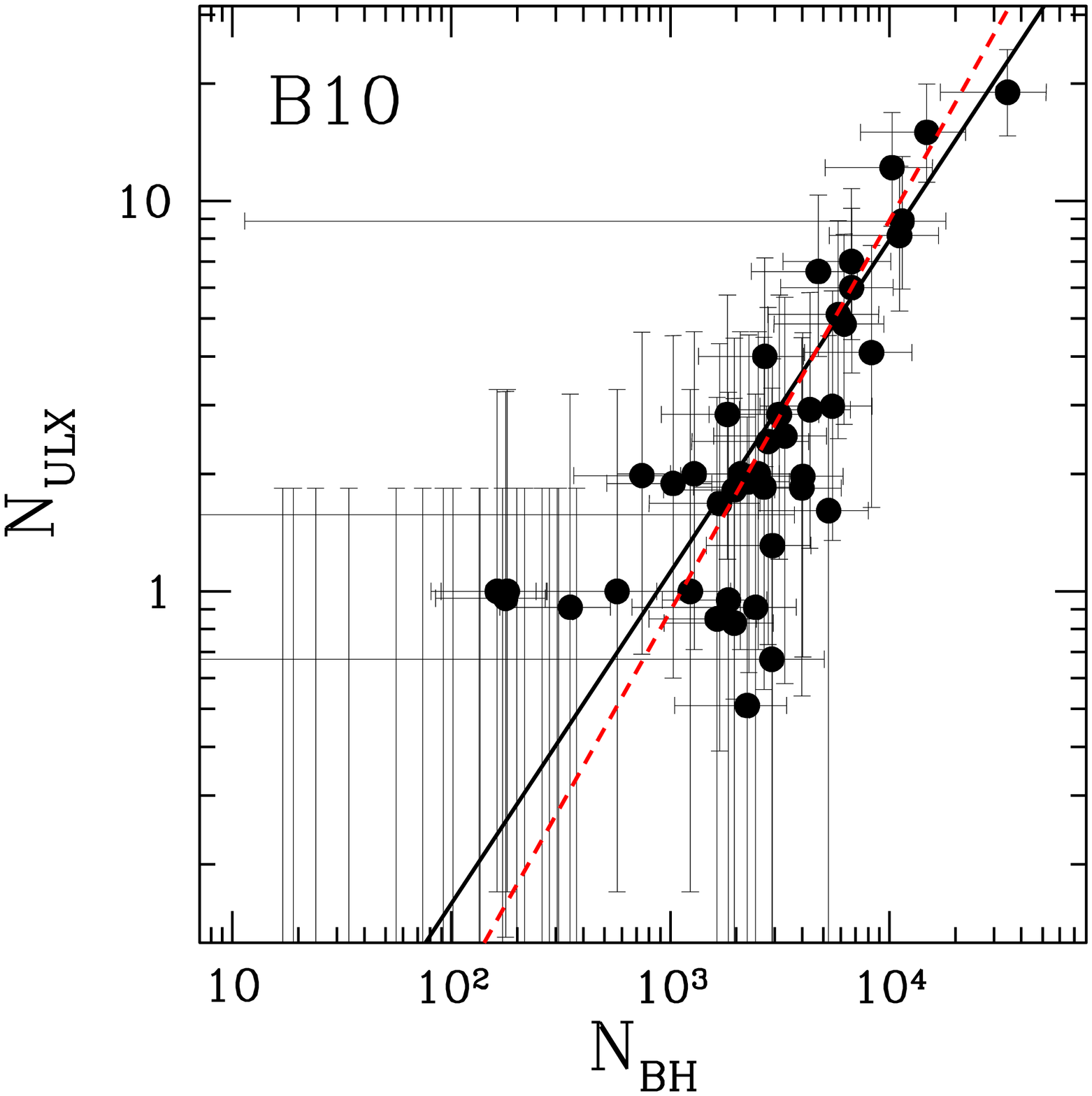,height=7cm} 
}}
\caption{\label{fig:figA1}
Number of expected massive BHs per galaxy, derived using the models from P98 (left-hand panel) and from B10 (right-hand panel), versus the number of observed ULXs per galaxy. The adopted IMF is a Salpeter IMF. The filled circles are the galaxies listed in Table~1.  The error bars on both the $x-$ and the $y-$ axis are $1-\sigma{}$ errors. The error bars on N$_{\rm BH}$ account for the uncertainty on the SFR and on the metallicity (see Section 2.3 for details). The solid lines are the  power-law fits, for the entire sample. The dashed lines (red on the web) are the power-law fits.
}
\end{figure*}
In this paper we assumed the Kroupa IMF (Kroupa 2001), a moderately top-heavy IMF. In this Appendix we show  how sensitive our model is to the adopted IMF, by using the Salpeter IMF (Salpeter 1995) to derive N$_{\rm BH}$. Table~A1 and Fig.~\ref{fig:figA1} show the results obtained adopting the Salpeter IMF. Basically, all the values of N$_{\rm BH}$ calculated with the Salpeter IMF are a factor of 2 lower than the ones obtained with the Kroupa IMF. However, the slope of the fit in equation~(\ref{eq:maingoal}) does not change significantly with a different IMF. Adopting the Salpeter IMF (Fig.~\ref{fig:figA1}), for the fits in equation~(\ref{eq:maingoal})  we find $\beta{}=0.80^{+0.16}_{-0.12}$ ($\beta{}=0.85^{+0.19}_{-0.13}$) and $\gamma{}= -2.12^{+0.43}_{-0.56}$ ($\gamma{}=-2.50^{+0.49}_{-0.70}$), corresponding to $\chi{}^2=8.7$ ($\chi{}^2=11.1$) with 62 dof, for the models by P98 (B10). As for the Kroupa IMF,  fixing $\beta{}=1$, we still obtain good values of $\chi{}^2$ ($\chi{}^2=10.0$ and 11.7 with 63 dof, for P98 and B10, respectively).
Thus, only the normalization of N$_{\rm BH}$ is affected by a different choice of the IMF. 
\begin{table*}
\begin{center}
\caption{Values of N$_{\rm BH}$ and $\epsilon_{\rm BH}$ for the galaxies listed in Table~1, assuming the models by P98 (first and second column) and by B10 (third and fourth column). We adopt a Salpeter IMF.} \leavevmode
\begin{tabular}[!h]{lllll}
\hline
Galaxy
& N$_{\rm BH}$ (P98)
& $\epsilon_{\rm BH}/10^{-4}$ (P98)
& N$_{\rm BH}$ (B10)
& $\epsilon_{\rm BH}/10^{-4}$ (B10)\\
\hline
The Cartwheel       & 20600$^{+10700}_{-10600}$      & $9.2^{+5.5}_{-5.2}$     &  34600$^{+17400}_{-17500}$  &  5.5$^{+3.2}_{-3.0}$\\
NGC~253             & 2970$^{+1560}_{-1590}$         & $10.0^{+11.1}_{-7.7}$     &  5490$^{+2830}_{-2930}$    &  $5.4^{+6.0}_{-4.1}$ \\
NGC~300             & 120$^{+61}_{-64}$              & $0^{+152.5}_{-0}$        &  216$^{+109}_{-111}$        &  $0^{+84.7}_{-0}$ \\
NGC~598       	    & 626$^{+348}_{-626}$          & $16.0^{+37.6}_{-16.0}$    &  1230$^{+650}_{-680}$     &  $8.1^{+19.1}_{-8.1}$\\
NGC~628             & 1400$^{+770}_{-760}$         & $13.7^{+20.2}_{-12.0}$     &  2680$^{+1420}_{-1400}$     &  $7.2^{+10.5}_{-6.2}$\\
NGC~1058	    & 185$^{+97}_{-100}$            & $49.2^{+126.5}_{-49.2}$  &  349$^{+184}_{-182}$        &  $26.1^{+67.0}_{-26.1}$ \\
NGC~1073            & 632$^{+356}_{-632}$          & $31.6^{+45.3}_{-31.6}$  &  1280$^{+680}_{-710}$   &  $15.6^{+22.1}_{-13.2}$ \\
NGC~1291	    & 1340$^{+670}_{-680}$         & $21.6^{+24.2}_{-16.5}$   &  1820$^{+920}_{-910}$     &  $15.9^{+17.9}_{-12.1}$\\
NGC~1313            & 1700$^{+860}_{-860}$         & $11.7^{+16.5}_{-9.6}$     &  2520$^{+1270}_{-1260}$     &  $7.9^{+11.2}_{-6.5}$\\
NGC~1365	    & 6190$^{+3150}_{-3280}$        & $13.2^{+9.5}_{-8.6}$     &  10280$^{+5420}_{-5200}$     &  $11.8^{+7.8}_{-7.0}$\\
IC~342              & 412$^{+222}_{-213}$            & $48.1^{+68.9}_{-40.2}$  &  741$^{+375}_{-379}$       &  $26.7^{+38.0}_{-22.2}$\\
NGC~1566            & 1740$^{+970}_{-1740}$         & $14.4^{+21.0}_{-14.4}$    &  3330$^{+1840}_{-1750}$    &  $7.5^{+11.0}_{-7.9}$ \\
NGC~1705            & 75$^{+38}_{-40}$              & $0^{+244.0}_{-0}$       &  135$^{+68}_{-70}$        &  $0^{+135.6}_{-0}$  \\
NGC~2366            & 91$^{+46}_{-47}$              & $0^{+201.1}_{-0}$        &  134$^{+68}_{-67}$        &  $0^{+136.6}_{-0}$  \\
NGC~2403	    & 312$^{+163}_{-166}$            & $32.0^{+75.3}_{-31.6}$  &  571$^{+294}_{-298}$       &  $17.5^{+41.1}_{-17.2}$\\
NGC~2442            & 0$^{+2580}_{-0}$               & $>2.6$                  &  2900$^{+2150}_{-2900}$     &  $2.3^{+10.0}_{-2.3}$\\
Holmberg~II         & 122$^{+62}_{-62}$            & $82.0^{+192.2}_{-79.7}$ &  180$^{+91}_{-90}$        &  $55.6^{+130.3}_{-53.9}$\\
NGC~2903            & 1190$^{+660}_{-670}$         & $16.0^{+23.7}_{-14.2}$    &  2280$^{+1210}_{-1190}$     &  $8.4^{+12.3}_{-7.3}$ \\
NGC~3031            & 2100$^{+1160}_{-1140}$         & $9.4^{+13.5}_{-8.0}$     &  4020$^{+2130}_{-2110}$     &  $4.9^{+7.0}_{-4.1}$ \\
NGC~3049            & 0$^{+0}_{-0}$                  & $-$                     &  0$^{+0}_{-0}$              &  $-$ \\
PGC~30819           & 54$^{+27}_{-28}$              & $0^{+338.9}_{-0}$       &  92$^{+47}_{-47}$          &  $0^{+198.9}_{-0}$\\
NGC~3310            & 1720$^{+930}_{-920}$         & $16.5^{+19.1}_{-13.2}$     &  3140$^{+1620}_{-1640}$      &  $9.0^{+10.4}_{-7.1}$ \\
NGC~3395/3396       & 3670$^{+1980}_{-1960}$         & $19.1^{+14.5}_{-12.4}$     &  6710$^{+3460}_{-3440}$     &  $10.4^{+7.8}_{-6.6}$ \\
PGC~35286           & 24$^{+12}_{-12}$               & $0^{+762.5}_{-0}$       & 34$^{+17}_{-17}$            &  $0^{+538.2}_{-0}$\\
PGC~35684           & 12$^{+6}_{-6}$               & $0^{+1525.0}_{-0}$       & 19$^{+10}_{-10}$            &  $0^{+963.2}_{-0}$\\
NGC~3738            & 31$^{+16}_{-17}$               & $0^{+590.3}_{-0}$       & 56$^{+28}_{-29}$           &  $0^{+326.8}_{-0}$\\
NGC~3972            & 152$^{+85}_{-152}$            & $0^{+120.4}_{-0}$       & 308$^{+164}_{-171}$         &  $0^{+59.4}_{-0}$\\
Antennae            & 11000$^{+5500}_{-5500}$      & $13.7^{+8.2}_{-7.8}$     & 14800$^{+7400}_{-7400}$   &  $10.1^{+6.1}_{-5.7}$\\
NGC~4144            & 41$^{+21}_{-22}$               & $0^{+446.3}_{-0}$       & 74$^{+37}_{-38}$           &  $0^{+247.3}_{-0}$\\
NGC~4214            & 105$^{+56}_{-55}$              & $0^{+174.3}_{-0}$        & 199$^{+101}_{-104}$         &  $0^{+92.0}_{-0}$\\
NGC~4236            & 94$^{+49}_{-50}$            & $0^{+194.7}_{-0}$        & 171$^{+87}_{-89}$         &  $0^{+107.0}_{-0}$\\
NGC~4248            & 12 $^{+7}_{-6}$              & $0^{+1525.0}_{-0}$       &  24$^{+12}_{-13}$           &  $0^{+762.5}_{-0}$\\
NGC~4254      	    & 2690$^{+1410}_{-1450}$         & $6.0^{+12.1}_{-6.0}$     &  5280$^{+2720}_{-2760}$    &  $3.0^{+6.1}_{-3.0}$\\
NGC~4258       	    & 1420$^{+790}_{-1420}$         & $17.0^{+22.9}_{-17.0}$    &  2790$^{+1480}_{-1540}$     &  $8.7^{+11.6}_{-8.0}$\\
NGC~4303      	    & 4750$^{+2480}_{-2530}$         & $8.6^{+9.2}_{-7.4}$     &  8280$^{+4370}_{-4190}$    &  $4.9^{+5.3}_{-4.1}$\\
NGC~4321            & 3060$^{+1680}_{-1650}$         & $16.7^{+16.0}_{-13.2}$     &  5840$^{+3090}_{-3060}$    &  $8.8^{+8.3}_{-6.8}$\\
NGC~4395	    & 97$^{+52}_{-52}$	     & $99.0^{+242.1}_{-99.0}$ &  177$^{+91}_{-92}$        &  $54.2^{+132.4}_{-54.2}$\\
NGC~4449            & 200$^{+105}_{-107}$            & $0^{+91.5}_{-0}$        &  373$^{+197}_{-195}$        &  $0^{+49.1}_{-0}$ \\
NGC~4485/4490       & 3370$^{+1760}_{-1800}$         & $14.5^{+12.6}_{-10.1}$     &  6220$^{+3200}_{-3250}$    &  $7.9^{+6.8}_{-5.4}$ \\
NGC~4501            & 0$^{+1550}_{-0}$               & $>8.3$    &  0$^{+3300}_{-0}$     &  $>3.9$\\
NGC~4559            & 1160$^{+610}_{-620}$         & $17.2^{+24.3}_{-14.4}$    &  2090$^{+1060}_{-1090}$     &  $9.6^{+13.5}_{-7.9}$\\
NGC~4631            & 2180$^{+1180}_{-1160}$         & $8.4^{+12.9}_{-7.7}$     &  3980$^{+2050}_{-2080}$     &  $4.6^{+7.0}_{-4.2}$ \\
NGC~4651            & 1020$^{+550}_{-530}$         & $8.1^{+22.9}_{-8.1}$    &  1960$^{+990}_{-1020}$     &  $4.2^{+11.9}_{-4.2}$\\
NGC~4656            & 1250$^{+630}_{-630}$         & $7.6^{+18.7}_{-7.6}$     &  1840$^{+920}_{-920}$     &  $5.2^{+12.7}_{-5.2}$\\
The Mice            & 3420$^{+1200}_{-3420}$         & $17.6^{+14.3}_{-17.6}$     &  6720$^{+3690}_{-3530}$    &  $8.9^{7.2}_{-5.9}$\\
NGC~4736            & 945$^{+481}_{-501}$          & $9.0^{+24.7}_{-9.0}$    &  1630$^{+840}_{-840}$     &  $5.2^{+14.3}_{-5.8}$\\
NGC~4861            & 638$^{+330}_{-328}$           & $29.6^{+44.0}_{-25.8}$  &  1030$^{+520}_{-520}$     &  $18.4^{+27.2}_{-15.9}$ \\
PGC~45561           &    0$^{+184}_{-0}$             & $0^{+99.5}_{-0}$        &  260$^{+145}_{-260}$        &  $0^{+70.4}_{-0}$\\
NGC~5033	    & 1990$^{+1010}_{-1000}$         & $6.8^{+14.0}_{-6.8}$     &  2930$^{+1470}_{-1470}$     &  $4.6^{+9.5}_{-4.6}$\\
NGC~5055      	    & 996$^{+553}_{-996}$         & $18.3^{+28.3}_{-18.3}$    &  1960$^{+1040}_{-1030}$     &  $9.3^{+14.3}_{-8.5}$\\
NGC~5194/5195       &    0$^{+8470}_{-0}$           & $>10.5$                  &  11400$^{+6700}_{-11400}$  &  $7.8^{+5.8}_{-7.8}$\\
NGC~5236            &    0$^{+1750}_{-0}$            & $>5.2$                  &  2450$^{+1310}_{-1780}$     &  $3.7^{+9.6}_{-3.7}$\\
NGC~5238            &    9$^{+5}_{-5}$             & $0{+2033.0}_{-0}$       &  17$^{+9}_{-9}$           &  $0^{+1076.0}_{-0}$\\
NGC~5408            & 109$^{+55}_{-55}$            & $91.7^{+215.2}_{-89.2}$ &  162$^{+82}_{-81}$        &  $61.7^{+144.8}_{-59.8}$\\
NGC~5457       	    & 2890$^{+1470}_{-1530}$        & $22.8^{+17.5}_{-15.1}$     &  4740$^{+2400}_{-2400}$     &  $13.9^{+10.6}_{-9.0}$\\
\noalign{\vspace{0.1cm}}
\hline
\end{tabular}
\footnotesize{\\{\it Table~A1 - continued.}}
\end{center}
\end{table*}
\begin{table*}
\begin{center}
\footnotesize{{\it Table~A1 - continued.}\\}
\begin{tabular}[!h]{lllll}
\hline
Galaxy
& N$_{\rm BH}$ (P98)
& $\epsilon_{\rm BH}/10^{-4}$ (P98)
& N$_{\rm BH}$ (B10)
& $\epsilon_{\rm BH}/10^{-4}$ (B10)\\
\hline
Circinus	    & 1820$^{+920}_{-920}$	     & $21.9^{+20.5}_{-15.2}$   &  2700$^{+1360}_{-1350}$     &  $14.8^{+13.9}_{-10.3}$\\
NGC~6946	    & 2270$^{+1250}_{-1230}$       & $12.9^{+14.6}_{-10.1}$    &  $4340^{+2300}_{-2270}$     & $6.7^{+7.6}_{-5.2}$\\
PGC~68618           & 1140$^{+600}_{-610}$       & $4.5^{+20.6}_{-4.5}$   &  $2240^{+1160}_{-1200}$     & $2.3^{+10.5}_{-2.3}$\\
NGC~7714/7715       & 6190$^{+3150}_{-3280}$      & $19.7^{+12.6}_{-12.0}$    &  $11100^{+5600}_{-5800}$  & $7.3^{+5.3}_{-4.7}$\\

NGC~7742            & 898$^{+489}_{-482}$         & $18.7^{+31.2}_{-18.6}$   &  $1680^{+880}_{-880}$     & $10.0^{+16.7}_{-9.9}$\\
Milky Way	    & 142$^{+79}_{-142}$          & $0^{+128.9}_{-0}$       &  $280^{+149}_{-147}$        & $0^{+65.4}_{-0}$\\
IC~10		    & 56$^{+30}_{-30}$            & $0^{+326.8}_{-0}$      &  $102^{+53}_{-53}$        & $0^{+179.4}_{-0}$\\
LMC                 & 159$^{+87}_{-86}$          & $0^{+115.1}_{-0}$       &  $304^{+161}_{-159}$        & $0^{+60.2}_{-0}$  \\
SMC                 & 154$^{+82}_{-78}$          & $0^{+118.8}_{-0}$       &  $260^{+130}_{-130}$        & $0^{+70.4}_{-0}$\\  
\noalign{\vspace{0.1cm}}
\hline
\end{tabular}
\end{center}
\end{table*}

\section{Data collection and elaboration}

In this section we provide details about how we collected and
calculated the values reported in Table~1, that represent the observational basis of this paper.

\subsection{Star Formation Rate}
As we already mentioned, the SFR reported in Tables~1 comes from
either ultra-violet (UV), H$\alpha$, far-infrared (FIR), or
radio measurements. If more then one independent measurement is
available, we generally average the corresponding SFRs, keeping into
account the properties of the galaxy, of the observational data, and
of calibrations.

In particular, for galaxies included in the UV catalogue by
Mu$\tilde{\rm n}$oz-Mateos et al. (2007), we derive
SFR$=2\,{}\pi{}r_{\alpha{}}^2\,{}\Sigma{}_{{\rm
SFR}0}\,{}\left[1-(1+r_{\rm tot}/r_{\alpha{}})\,{}\exp{(-r_{\rm
tot}/r_{\alpha{}})}\right]$, where $\Sigma{}_{{\rm SFR}0}$ is the
central SFR surface density, $r_{\alpha{}}$ is a characteristic
length-scale and $r_{\rm tot}$ is the external radius of the galaxy
(from the integration of equation B1a in Mu$\tilde{\rm n}$oz-Mateos et
al. 2007).  When the total H$\alpha{}$ luminosities are available, we
adopt the correlation in Kennicutt (1998; hereafter K98), i.e. SFR$=L({\rm
H}\alpha{})/1.26\times{}10^{41}$ erg s$^{-1}$ $[{\rm M}_\odot{}$
yr$^{-1}]$.  Similarly, when the FIR luminosity is available, we apply
the equation SFR$=L({\rm FIR})/2.2\times{}10^{43}$ erg s$^{-1}$ $[{\rm
M}_\odot{}$ yr$^{-1}]$ in K98.  When a radio measurement
(at 1.4 GHz) is available, we then use equation~(6) of Bell (2003).
Finally, if a galaxy belongs to the sample by Grimm et al. (2003), we
 generally use their adopted values.

\subsection{Metallicity}
The existence of a metallicity measurement is the most restrictive
among the criteria for inclusion in our sample, as metallicities
are often unavailable.

For most galaxies in the sample we use measurements of oxygen
abundance (based on oxygen line intensities from HII regions) as
proxies for metallicity. Instead, when the spectra of HII regions are
unavailable, we use X-ray metallicity estimates (from papers on X-ray measurements of ULXs), although they are much less
accurate. In the rest of this subsection we provide extra details
about the measurements from HII regions.

\subsubsection{Conversion of line intensities to metallicities}
When line intensity measurements include the weak $\lambda$=4363\AA\
OIII line (e.g. in the case of HII regions of the Cartwheel galaxy
from Fosbury \& Hawarden 1979), we simply use the oxygen abundances
reported in the literature: this is possible because the measurement
of the 4363\AA\ line greatly simplifies the conversion of line
intensities into abundances, and the conversion procedure did not
change much since it was first established (see e.g. Pagel
et~al. 1992; and the discussion in PT05).
Instead, when only the most intense OII ($\lambda$=3727\AA) and
OIII ($\lambda$=4959,5007\AA) lines are observed,
the oxygen abundances in
different papers are often based on different calibrations (see the
discussion in P01, and in PT05), that lead to significant offsets in
the results. For this reason, we went back to the observational data
and applied the PT05 calibration whenever this was possible. We made
an exception for galaxies where the metallicity value in the
literature had been calculated with the P01 calibration (such as the
ones in the compilation by Pilyugin, V\'ilchez \&{} Contini 2004;
hereafter PVC04); this is because the two calibrations are quite close
to each other (the PT05 calibration is essentially a revision of the
P01 calibration), and the difference (typically amounting to
$\sim 0.05$ dex for single HII regions; see the top panel of Fig.~7
from PT05) should not have a large effect.

We note that both the PT05, and the P01 calibrations generally provide
significantly lower abundances with respect to previous calibrations
(e.g. Edmunds \&{} Pagel 1984), but are currently considered the most
accurate (see e.g. Kennicutt, Bresolin \&{} Garnett 2003, who actually
suggest that abundances might be even lower).

When the metallicity gradient is available for a given galaxy, we list
in Table~1 and adopt in our calculations the value of the
metallicity at $r=0.73\,{}R_{25}$, where $R_{25}$ is the isophotal or
Holmberg radius. In fact, $r=0.73\,{}R_{25}$ is the average distance of
the observed ULXs from the centre of the host galaxy, in a sample of
spiral galaxies (fig.~15 in Liu et al. 2006; the averaging process
took contamination into account).

The oxygen abundance is usually derived in terms of $12+{\rm log}({\rm
O/H})$; this can be easily converted to the units used in Table~1
(where the metallicity $Z$ is expressed as a fraction of solar
metallicity $Z_\odot{}$), by assuming
 that $12+{\rm log}({\rm O/H})_\odot=8.92$, which corresponds to
 $Z_\odot{}=0.02$  (the value commonly used in stellar evolution
 studies).

\subsection{Distance}
 For the 52 galaxies in the LB05
catalogue we use the distance reported there. For the MW
we simply indicate the distance to the Galactic centre ($8.5\,{\rm
  kpc}$). For the  remaining 11 galaxies we use the distances reported by
the NASA Extragalactic Database (NED). In the 9 cases where it is
possible, we take the mean redshift-independent distance; for the
remaining 2 galaxies we derive distances from the redshift, assuming
$H_0=73\,{\rm km\,s^{-1}\,Mpc^{-1}}$.

\subsection{Contaminating sources}

\subsubsection{Further required data}
Estimating the number of contaminating sources requires the use of several quantities that are not reported in Table~1: the galaxy
angular size (major and minor axis $R_{25}$ and $r_{25}$ of the
$25\;{\rm mag\, arcsec^{-2}}$ isophote), the sky area $A_{\rm obs}$
actually covered by the X-ray observations, and the total galactic
HI column density ($N_{\rm H}$).

For galaxies in the LB05 catalogue, we use the $R_{25}$, $r_{25}$ and
$N_{\rm H}$ values reported there. For the others, we adopt the references quoted in LB05: the RC3 catalogue (de Vaucouleurs
et~al. 1991) for angular sizes, and the paper of Dickey \&{} Lockman (1990; hereafter DL90) for the weighted average of $N_{\rm H}$. We make an
exception for the MW and the Magellanic Clouds, where such quantities
do not really matter, since the number density of the expected
contaminating sources is extremely low, and there are no ULXs inside
such galaxies.

The sky area $A_{\rm obs}$ was taken from the papers reporting the
X-ray observations. When no specific information is provided, we
assume an area of $314\,{\rm arcmin^2}$ if the data were taken with
ROSAT (this corresponds to the area within $10'$ from the instrument's
axis, where the sensitivity is reasonably constant and close to the
maximum value; see LB05), or an area of $70.6\,{\rm arcmin^2}$
(i.e. the entire $8.4'\times8.4'$ field of view of the S3 detector)
if the data were taken with {\it Chandra}.

\subsubsection{Contamination estimate}

In order to estimate the number of background or foreground
contaminating sources, we followed a procedure very similar to the one
reported by Liu et~al. (2006).

Since the bulk of the galaxies in our sample was observed with ROSAT,
we used the log(N)-log(S) reported in Hasinger et~al. (1998).  We
converted the minimum luminosity of an ULX, $L_{\rm lim}=10^{39}\,{\rm
erg\,s^{-1}}$, into the apparent flux limit $F_{\rm lim}=L_{\rm
lim}/(4\pi D^2)$; then, we converted $F_{\rm lim}$ into another flux
$S_{\rm lim}$ that can be used in the Hasinger et~al. (1998)
log(N)-log(S) relation. Such conversion takes into account (i) the
different assumptions on the observed band (the Hasinger et~al. 1998
flux refers to the 0.5--2.4 keV band, whereas most references for ULXs
provide luminosities in other bands - e.g. 0.3--8.0 keV band); (ii) the
different assumption over spectral slopes (Hasinger et~al. 1998
assumes that all sources have photon index 2; whereas here we assume a
photon index 1.7, as done by LB05); (iii) the absorption from
the galactic $N_{\rm H}$. In most cases, such conversion amounts to a
reduction by a factor of 2--3 (the exact value depending on the band used
for the X-ray observations, and to a lesser extent on the galactic
$N_{\rm H}$ for each specific galaxy) of $F_{\rm lim}$. Finally, we
integrate the Hasinger et~al. (1998) differential log(N)-log(S), and
find the expected surface number density $q$ of contaminating sources
(i.e. of sources with apparent flux larger than $S_{\rm lim}$).

In order to determine the expected contamination in a specific galaxy,
we combined $q$ with the size of the galaxy (and of the observed area), taking into account also the radial distribution of ULXs. The
exact procedure is the following. First of all, we calculate the
observed area with deprojected distance $\le R_{25}$ from the centre
 $A_1=\min(A_{\rm obs},\pi R_{25} r_{25})$, and the observed area
with deprojected distance between $R_{25}$ and $2R_{25}$,
$A_2=\min(A_{\rm obs}-A_1,3 \pi R_{25} r_{25})$. This assumes that the
field of the X-ray observation was centred on the host galaxy, and
that such field is circular. Then, we obtain the number of expected
contaminating sources in the two regions, $Q_1 = A_1 q$ and $Q_2 = A_2
q$. These must be compared to the numbers ${\rm N}_1$ and ${\rm N}_2$ 
of observed ULXs in the corresponding areas (obviously, 
${\rm N}_1+{\rm N}_2={\rm N}_{{\rm ULX,\,{}raw}}$), since the number of 
contaminating sources cannot be larger
than the number of observed sources. Then, we estimate the actual
number of contaminating sources to be
\begin{equation}
Q = \min(Q_1,{\rm N}_1) + \min(Q_2,{\rm N}_2).
\end{equation}
This expression automatically sets the number of
contaminating sources at 0 for galaxies where no ULX was observed.

For the 10 galaxies where the original X-ray observations provide a
value of the number of ULXs that was already cleared from
contaminating sources (e.g. because one or more of the candidate ULXs
in a galaxy have been identified as background objects), we simply
assume that N$_{{\rm ULX,\,{}raw}}={\rm N}_{\rm ULX}$, i.e. that
$Q=0$.

The total number of expected contaminating sources in the  54
galaxies where we apply our estimates  is 14.60, whereas the sum
of N$_{{\rm ULX,\,{}raw}}$ in such galaxies  is 112; the
contamination fraction is  then $\simeq$0.13, which is smaller
than those found in previous papers ($25-44$ per cent, Swartz et
al. 2004; Ptak \&{} Colbert 2004; Liu et al. 2006; L\'opez-Corredoira
\&{} Guti\'errez 2006). There are two reasons for this
discrepancy. First, we do not consider elliptical galaxies, where the
contamination is stronger: for example, Liu et al. (2006) estimate a
contamination fraction $\sim1$ for early type galaxies, whereas for
late-type galaxy such fraction is only $\sim0.2$. Second, our sample
includes a smaller fraction of distant galaxies than most ULXs
catalogues (for example, in the LB05 catalogue the fraction of
galaxies with distances $\ge20\,{\rm Mpc}$ is $\sim0.4$, whereas in
our sample it is $\lesssim 0.1$): this is relevant because distant
galaxies tend to have higher contamination levels than nearby
galaxies\footnote{ The distance D affects of the contamination $Q$
through the angular surface of a galaxy, and the corresponding limit
flux $S_{\rm lim}$. Both scale as D$^{-2}$. However, the Hasinger
et~al. (1998) $\log({\rm N})-\log({\rm S})$ implies that for
D$\lesssim 30\,{\rm Mpc}$ the decrease in $S_{\rm lim}$ produces a
steep increase in the surface density of contaminating sources
($q\propto {\rm D}^{3.4}$), and $Q\propto {\rm D}^{1.4}$; instead, for
sources at large distances the two dependencies almost cancel out, and
$Q\propto {\rm D}^{-0.1}$.}.

\subsection{A possible contamination from Population II ULXs}
 ULXs in elliptical galaxies might be connected to the minor (but non-negligible) fraction of ULXs that is claimed (e.g. Colbert et~al. 2004; Brassington et~al. 2005) to be associated with old stellar populations (Population II ULXs). Since old populations are present also in spiral galaxies, Population II ULXs might be present also in our sample.  It is quite difficult to estimate their number. 

Liu et~al. (2006) estimate that the number of ULXs in early-type and elliptical galaxies can be completely explained by contaminating sources (i.e. without requiring Population II ULXs): if so, the number of ULXs associated to old stellar populations should be very low (or even negligible) also in spiral/irregular galaxies. On the other hand, Colbert et~al. (2004) suggest that Population II ULXs represent a fraction $\sim0.2$ of ULXs in spirals. However, the results by Colbert et~al. (2004) and by Liu et~al. (2006) are not necessarily in conflict. In fact, 1-$\sigma$ upper limits for early-type galaxies\footnote{In their Section~3.2 and Tables~1 and 2, Liu et~al. (2006) estimate the occurrence frequencies of ULXs. They find $0.72\pm0.11$ ULXs per spiral galaxy, $0.02\pm0.11$ ULXs per early-type galaxy, and $-0.15\pm0.13$ ULXs per elliptical galaxy. The ratio is then $0.03\pm0.15$ if we consider early-type galaxies, and $-0.21\pm0.19$ if we consider ellipticals. The ratio for early-types can be taken as an upper limit on the fraction of Population II ULXs in spirals, since the old population of early-type/elliptical galaxies of the Liu et~al.(2006) sample is typically larger than the old population in the spiral/irregular galaxies of the same sample.} from Liu et~al. (2006) are not far from the Colbert et~al. (2004) results. Since the fraction of  population~II ULXs in spiral galaxies is small for any of these estimates, we neglect their contribution to our ULX sample.

\subsection{References and details for the single galaxies}

 Here we give the detailed references for the data used to compile Table~1, grouped by galaxy.

{\bf The Cartwheel (ESO~350--G 040)}: average SFR from Appleton \&{} Marston (1997, H$\alpha{}$ data) and from Mayya et al. (2005, radio measurement); $Z$ from Fosbury \& Hawarden (1977), spectra of HII regions in the outer ring; X-ray sources from Wolter \& Trinchieri (2004). This galaxy is not in the LB05 catalogue. Distance from the NASA Extragalactic Database (hereafter NED), assuming $H_0=73\,{\rm km\,{}s^{-1} Mpc^{-1}}$ and $V({\rm Local Group})=9048\,{\rm km\,{}s^{-1}}$. Angular size and axis ratio from the RC3 catalogue; $N_{\rm H}$ from DL90.

{\bf NGC~253}: SFR from Grimm, Gilfanov \&{} Sunyaev (2003); $Z$ from  Webster \&{} Smith (1983), that reports the measurements of line intensities for OII$[\lambda{}3727+\lambda{}3729]$ and OIII$[\lambda{}4959+\lambda{}5007]$, from which we derive the metallicity with the formula by P01; X-ray  from LM05.

{\bf NGC~300}: SFR from Helou  et al. (2004), H$\alpha{}$ measurement, using the correlation by K98; metallicity gradient from PVC04; X-ray from LB05.

{\bf NGC~598 (M~33)}: SFR from ultra-violet (Mu$\tilde{\rm n}$oz-Mateos et al. 2007); metallicity gradient from spectra of HII regions (PVC04); X-ray from LM05. This galaxy is not in the LB05 catalogue. Distance from NED (mean redshift-independent distance). Angular size and axis ratio from the RC3 catalogue; $N_{\rm H}$ from DL90.

{\bf NGC~628 (M~74)}: SFR from Grimm et al. (2003); metallicity gradient from spectra of HII regions (Pilyugin et al. 2002); X-ray from LM05. This galaxy is not in the LB05 catalogue. Distance from NED (mean redshift-independent distance). Angular size and axis ratio from the RC3 catalogue; $N_{\rm H}$ from DL90.

{\bf NGC~1058}: SFR from H$\alpha{}$ (Kennicutt\&{} Kent 1983), using the correlation by K98; metallicity gradient from Moustakas \&{} Kennicutt (2006a), based on the PT05 calibration; X-ray from LM05. This galaxy is not in the LB05 catalogue. Distance from NED (mean redshift-independent distance). Angular size and axis ratio from the RC3 catalogue; $N_{\rm H}$ from DL90.

{\bf NGC~1073}: SFR from  H$\alpha{}$ (Martin \&{} Friedli 1997), using the correlation by K98; metallicity gradient from spectra of HII regions (Dors \&{} Copetti 2005); X-ray from LB05. 

{\bf NGC~1291}: SFR from ultra-violet (Mu$\tilde{\rm n}$oz-Mateos et al. 2007);  $Z$ from X-ray measurements (Irwin, Sarazin \&{} Bregman 2002); X-ray from LM05. 

{\bf NGC~1313}: SFR from H$\alpha{}$ (Ryder \& Dopita 1994); $Z$ from spectra of HII regions (P01); X-ray from Colbert et al. (1995).

{\bf NGC~1365}: average SFR from  mid- and far-infrared measurements (Lonsdale \&{} Helou 1985; Strateva \&{} Komossa 2009a, 2009b); metallicity gradient from spectra of HII regions (PVC04);  X-ray from Strateva \&{} Komossa (2009a).  

{\bf IC~342 (PGC~13826)}: SFR from Grimm et al. (2003); metallicity gradient from spectra of HII regions (PVC04); X-ray from LM05. 

{\bf NGC~1566}:  SFR from far-infrared (SFR$_{\rm FIR}=3.24\,{}{\rm M}_{\odot}\,{}{\rm yr}^{-1}$, Bell \&{} Kennicutt 2001, using the correlation by K98); $Z$ from spectra of HII regions (Hawley \&{} Phillips 1980); X-ray from LB05.

{\bf NGC~1705}: average SFR from two different H$\alpha$ measurements (SFR$_{{\rm H}\alpha{}}=0.063\,{}{\rm M}_{\odot}\,{}{\rm yr}^{-1}$, Gil de Paz, Madore \&{} Pevunova 2003; SFR$_{{\rm H}\alpha{}}=0.112\,{}{\rm M}_{\odot}\,{}{\rm yr}^{-1}$, Hunter \&{} Elmegreen 2004), using the correlation by K98;	$Z$ from Lee \&{} Skillman (2004), spectra of HII regions; X-ray from LB05.
  
{\bf NGC~2366}: average SFR from H$\alpha{}$ (SFR$_{{\rm H}\alpha{}}=0.125\,{}{\rm M}_{\odot}\,{}{\rm yr}^{-1}$, Hunter \&{} Elmegreen 2004, using the calibration by K98, 8$-$1000 $\mu{}$m (SFR$_{{\rm IR}}=0.03\,{}{\rm M}_{\odot}\,{}{\rm yr}^{-1}$, Hopkins, Schulte-Ladbeck \&{} Drozdovsky 2002, using the calibration by K98 and radio (SFR$_{1.4{\rm GHz}}=0.069\,{}{\rm M}_{\odot}\,{}{\rm yr}^{-1}$, Condon, Cotton \&{} Broderick 2002, using the calibration by Bell 2003); $Z$ from spectra of HII regions (PVC04); X-ray from LB05.

{\bf NGC~2403}: SFR from Gilfanov, Grimm \&{} Sunyaev (2004a); metallicity gradient from spectra of HII regions (PVC04); X-ray from LM05. 

{\bf NGC~2442}: average SFR from  ultra-violet (SFR$_{\rm UV}=3.86\,{}{\rm M}_{\odot}\,{}{\rm yr}^{-1}$, Mu$\tilde{\rm n}$oz-Mateos et al. 2007) and H$\alpha{}$ (SFR$_{{\rm H}\alpha{}}=5.26\,{}{\rm M}_{\odot}\,{}{\rm yr}^{-1}$, Helmboldt et al 2004, using the calibration by K98); metallicity gradient from spectra of HII regions (PVC04); X-ray from LB05.

{\bf Holmberg~II (Arp~268, PGC~23324)}: SFR from H$\alpha{}$ (Walter et al. 2007), adopting the relation in K98; $Z$ from spectra of HII regions (Walter et al. 2007); X-ray from Dewangan et al. (2004). 

{\bf NGC 2903}:  average SFR from far-infrared (SFR$_{\rm FIR}=1.8\,{}{\rm M}_{\odot}\,{}{\rm yr}^{-1}$, Ranalli et al. 2003, using the calibration by K98; SFR$_{\rm FIR}=1.5\,{}{\rm M}_{\odot}\,{}{\rm yr}^{-1}$, Bell 2003, using the calibration by K98) and radio (SFR$_{1.4{\rm GHz}}=2.6\,{}{\rm M}_{\odot}\,{}{\rm yr}^{-1}$, Ranalli et al. 2003, using the calibration by Bell 2003; SFR$_{1.4{\rm GHz}}=1.6\,{}{\rm M}_{\odot}\,{}{\rm yr}^{-1}$, Bell 2003);  metallicity gradient from spectra of HII regions (Pilyugin, Thuan \&{} V\'ilchez 2006); X-ray from LB05.

{\bf NGC~3031 (M~81)}: SFR from ultra-violet (Mu$\tilde{\rm n}$oz-Mateos et al. 2007); metallicity gradient from spectra of HII regions (PVC04); X-ray from LM05. 

{\bf NGC~3049}: SFR from radio (Bell 2003); average $Z$ from  Guseva, Izotov \&{} Thuan (2000) and Kehrig, Telles \&{} Cuisinier (2004); X-ray from LB05.

{\bf PGC~30819 (IC~2574, UGC~05666)}: average SFR from H$\alpha{}$ (SFR$_{{\rm H}\alpha{}}=0.076\,{}{\rm M}_{\odot}\,{}{\rm yr}^{-1}$, Miller \&{} Hodge 1994, using the calibration by K98) and radio (SFR$_{1.4{\rm GHz}}=0.039\,{}{\rm M}_{\odot}\,{}{\rm yr}^{-1}$, Condon et al. 2002, using the calibration by Bell 2003); $Z$ from spectra of HII regions (Miller \&{} Hodge 1996); X-ray from LB05.

{\bf NGC~3310 (Arp~217)}: SFR from Grimm et al. (2003); $Z$ from spectra of HII regions  (Pastoriza et al. 1993), metallicity from spectra of HII regions; X-ray from LM05. 

{\bf NGC~3395/3396 (Arp~270)}: SFR from radio (Condon et al. 2002), using the calibration by Bell (2003); $Z$ from integrated spectrum (Kennicutt 1992); X-ray from Brassington et~al. (2005). 

{\bf PGC~35286 (UGC~06456)}: SFR from H$\alpha{}$ (Hunter \&{} Elmegreen 2004), using the calibration by K98; $Z$ from spectra of HII regions (Pilyugin 2001b);   X-ray from LB05.

{\bf PGC~35684 (UGC~06541, Mkn~178)}: SFR from H$\alpha{}$ (Hunter \&{} Elmegreen 2004), using the calibration by K98; $Z$ from spectra of HII regions (Pilyugin 2001b); X-ray from LB05.

{\bf NGC~3738 (Arp~234)}: average SFR from H$\alpha{}$ (SFR$_{{\rm H}\alpha{}}=0.047\,{}{\rm M}_{\odot}\,{}{\rm yr}^{-1}$, Hunter \&{} Elmegreen 2004) and radio (SFR$_{1.4{\rm GHz}}=0.029\,{}{\rm M}_{\odot}\,{}{\rm yr}^{-1}$, Condon et al. 2002, using the calibration by Bell 2003); $Z$ from spectra of HII regions (Martin 1997); X-ray from LB05.

{\bf NGC~3972}:  average SFR from H$\alpha{}$ (SFR$_{{\rm H}\alpha{}}=0.15\,{}{\rm M}_{\odot}\,{}{\rm yr}^{-1}$, Moustakas \&{} Kennicutt 2006b, using the calibration by K98), far-infrared (SFR$_{\rm FIR}=0.38\,{}{\rm M}_{\odot}\,{}{\rm yr}^{-1}$, Bell 2003, using the calibration by K98) and radio (SFR$_{1.4{\rm GHz}}=0.36\,{}{\rm M}_{\odot}\,{}{\rm yr}^{-1}$, Bell 2003); $Z$ from spectra of HII regions (PT05); X-ray from LB05.

{\bf The Antennae (NGC~4038/4039, Arp~244)}: SFR from Grimm et al. (2003);  X-ray estimate of $Z$ (Fabbiano et al. 2004); X-ray from Fabbiano, Zezas \& Murray (2001), spectral fits from LM05. The {\it Chandra} observation was analyzed only in a central area of $\sim{}5.9$ arcmin$^2$. This galaxy is not in the LB05 catalogue. Distance from NED (mean redshift-independent distance). Angular sizes and axis ratios from the RC3 catalogue; $N_{\rm H}$ from DL90.

{\bf NGC~4144}: average SFR from H$\alpha{}$ (SFR$_{{\rm H}\alpha{}}=0.05\,{}{\rm M}_{\odot}\,{}{\rm yr}^{-1}$, Moustakas \&{} Kennicutt 2006b, using the calibration by K98) and radio (SFR$_{1.4{\rm GHz}}=0.05\,{}{\rm M}_{\odot}\,{}{\rm yr}^{-1}$, Condon et al. 2002, with the calibration by Bell 2003);$Z$ from spectra of HII regions (PT05); X-ray from LB05.

{\bf NGC~4214}: average SFR from H$\alpha{}$ (SFR$_{{\rm H}\alpha{}}=0.18\,{}{\rm M}_{\odot}\,{}{\rm yr}^{-1}$, Hunter \&{} Elmegreen 2004; SFR$_{{\rm H}\alpha{}}=0.10\,{}{\rm M}_{\odot}\,{}{\rm yr}^{-1}$, Schmitt et al. 2006, using the calibration by K98), far-infrared (SFR$_{{\rm FIR}}=0.15\,{}{\rm M}_{\odot}\,{}{\rm yr}^{-1}$, Bell 2003; SFR$_{{\rm FIR}}=0.11\,{}{\rm M}_{\odot}\,{}{\rm yr}^{-1}$, Schmitt et al. 2006, using the calibration by K98) and radio (SFR$_{1.4{\rm GHz}}=0.13\,{}{\rm M}_{\odot}\,{}{\rm yr}^{-1}$, Bell 2003); $Z$ from spectra of HII regions (PT05); X-ray from LB05.

{\bf NGC~4236}: SFR from radio (Condon et al. 2002, using the calibration by Bell 2003);  $Z$ from spectra of HII regions (Vigroux, Stasi\'nska \&{} Comte 1987);   X-ray from LB05.

{\bf NGC~4248}: SFR from Moustakas \&{} Kennicutt (2006b); $Z$ from spectra of HII regions (PT05); X-ray from LB05.

{\bf NGC~4254 (M~99)}:  average SFR from H$\alpha{}$ (SFR$_{{\rm H}\alpha{}}=3.9\,{}{\rm M}_{\odot}\,{}{\rm yr}^{-1}$, Kennicutt, Tamblyn \&{} Congdon 1994; SFR$_{{\rm H}\alpha{}}=5.4\,{}{\rm M}_{\odot}\,{}{\rm yr}^{-1}$, Buat et al. 2002; SFR$_{{\rm H}\alpha{}}=3.2\,{}{\rm M}_{\odot}\,{}{\rm yr}^{-1}$, Moustakas \&{} Kennicutt 2006b) and far-infrared (SFR$_{{\rm FIR}}=3.48\,{}{\rm M}_{\odot}\,{}{\rm yr}^{-1}$, Buat et al. 2002); metallicity gradient from spectra of HII regions (Moustakas \&{} Kennicutt 2006a); X-ray from LB05. 

{\bf NGC~4258 (M~106)}: SFR from radio (Condon et al. 2002, using the calibration by Bell 2003); metallicity gradient from spectra of HII regions (PVC04); X-ray from LB05. 

{\bf NGC~4303 (M~61)}: average SFR from ultra-violet (SFR$_{\rm UV}=8\,{}{\rm M}_{\odot}\,{}{\rm yr}^{-1}$, Mu$\tilde{\rm n}$oz-Mateos et al. 2007) and from H$\alpha{}$ (SFR$_{{\rm H}_\alpha{}}=3.6\,{}{\rm M}_{\odot}\,{}{\rm yr}^{-1}$, Moustakas \&{} Kennicutt 2006b); metallicity gradient from spectra of HII regions (Moustakas \&{} Kennicutt 2006a);  X-ray from LB05 and from Tsch\"oke, Hensler \&{} Junkes (2000), whose sources B and F are not present in LB05.

{\bf NGC~4321 (M~100)}: SFR from Grimm et al. (2003);	metallicity gradient from spectra of HII regions (Moustakas \&{} Kennicutt 2006a); X-ray from Kaaret (2001) and from LB05. 

{\bf NGC~4395}: SFR from ultra-violet (Mu$\tilde{\rm n}$oz-Mateos et al. 2007); metallicity gradient from spectra of HII regions (PVC04); X-ray from LM05. 

{\bf NGC~4449}: average SFR from far-infrared (SFR$_{{\rm FIR}}=0.22\,{}{\rm M}_{\odot}\,{}{\rm yr}^{-1}$, Bell 2003; SFR$_{{\rm FIR}}=0.21\,{}{\rm M}_{\odot}\,{}{\rm yr}^{-1}$, Ranalli et al. 2003, using the calibration by K98) and radio (SFR$_{1.4{\rm GHz}}=0.42\,{}{\rm M}_{\odot}\,{}{\rm yr}^{-1}$, Bell 2003); $Z$ from spectra of HII regions (Skillman, Kennicutt \&{} Hodge 1989); X-ray from LB05.

{\bf NGC~4485/4490 (Arp~269)}: average SFR from H$\alpha{}$ (SFR$_{{\rm H}_\alpha{}}=5.49\,{}{\rm M}_{\odot}\,{}{\rm yr}^{-1}$, Kennicutt et al. 1987, assuming the calibration by K98 and an extinction of $\sim{}1$ mag,  Clemens, Alexander \&{} Green  1999);  metallicity gradient from spectra of HII regions (Pilyugin \&{} Thuan 2007);  X-ray from LM05. 

{\bf NGC~4501 (M~88)}: SFR from radio measurements by Condon et al. (2002); other measurements indicate both lower values ($1.4\,{\rm M}_\odot{}$ yr$^{-1}$, from H$\alpha$ photometry reported in Gavazzi et~al. 2002), and higher values ($6.0\,{\rm M}_\odot{}$ yr$^{-1}$, from FIR measurements reported in Spinoglio et~al. 2002); metallicity gradient from spectra of HII regions (Pilyugin et al. 2002); X-ray from LM05.

{\bf NGC~4559}: SFR from radio (Condon et al. 2002), using the calibration by Bell (2003); metallicity gradient from spectra of HII regions (PVC04); X-ray from Cropper et al. (2004).

{\bf NGC~4631 (Arp~281)}: SFR from far-infrared (Ranalli et al. 2003, using the correlation by K98);  $Z$ from spectra of HII regions (Pilyugin, Izotova \&{} Sholudchenko 2008); X-ray from LM05. 

{\bf NGC~4651 (Arp~189)}: average SFR from H$\alpha{}$ (SFR$_{{\rm H}_\alpha{}}=1.2\,{}{\rm M}_{\odot}\,{}{\rm yr}^{-1}$, Hirashita et al. 2003, using the correlation by K98), far-infrared (SFR$_{{\rm FIR}}=1.6\,{}{\rm M}_{\odot}\,{}{\rm yr}^{-1}$, Bell 2003, using the calibration by K98) and radio (SFR$_{1.4{\rm GHz}}=1.3\,{}{\rm M}_{\odot}\,{}{\rm yr}^{-1}$, Bell 2003); metallicity gradient from spectra of HII regions (Moustakas \&{} Kennicutt 2006a); X-ray from LB05.

{\bf NGC~4656}: average SFR from H$\alpha{}$ (SFR$_{{\rm H}_\alpha{}}=0.54\,{}{\rm M}_{\odot}\,{}{\rm yr}^{-1}$, Moustakas \&{} Kennicutt 2006b) and radio (SFR$_{1.4{\rm GHz}}=1.35\,{}{\rm M}_{\odot}\,{}{\rm yr}^{-1}$, Condon et al. 2002, using the calibration by Bell 2003);  $Z$ from spectra of HII regions (Pilyugin, Izotova \&{} Sholudchenko 2008);  X-ray from LB05.

{\bf The Mice (NGC~4676, Arp~242)}: SFR from far-infrared (Brassington, Ponman \&{} Read 2007, using the correlation by K98); X-ray estimate of $Z$ (Read 2003); X-ray  from Read (2003); but these observations are complete only for objects with $L_{\rm X}\gtrsim 3\times10^{39}\,{\rm erg\,s^{-1}}$; so, it is quite likely that the actual number of ULX is larger than what we report in Table~1 (6). This galaxy is not in the LB05 catalogue. Distance from NED (assuming $H_0=73\,{\rm km\,{}s^{-1} Mpc^{-1}}$ and $V({\rm Local Group})=6602\,{\rm km\,{}s^{-1}}$). Angular sizes and axis ratios from the RC3 catalogue; $N_{\rm H}$ from DL90.

{\bf NGC~4736 (M~94)}: SFR from Grimm et al. (2003); metallicity gradient from Moustakas \&{} Kennicutt (2006a);  X-ray from LB05.

{\bf NGC~4861 (Arp~266)}: SFR from H$\alpha{}$ (Schmitt et al. 2006), adopting the relation in  K98; $Z$ from spectra of HII regions (Izotov, Thuan \&{} Lipovetsky 1997); X-ray from LM05. 

{\bf PGC~45561 (UGC~08231)}: SFR from H$\alpha{}$ (Kewley et al. 2002), adopting the relation in K98; $Z$ from spectra of HII regions (Kewley, Jansen \&{} Geller 2005);  X-ray from LB05.

{\bf NGC~5033}: SFR from H$\alpha{}$ (Kennicutt \&{} Kent 1983), using the correlation by K98; metallicity gradient from spectra of HII regions (PVC04); X-ray from LM05.

{\bf NGC~5055 (M~63)}: SFR from radio (Bell 2003); metallicity gradient from spectra of HII regions (PVC04); X-ray from LM05.

{\bf NGC~5194/5195 (M~51, Arp~85)}:  SFR from ultra-violet (Mu$\tilde{\rm n}$oz-Mateos et al. 2007); metallicity gradient from spectra of HII regions with T$_e$ method (PT05);  X-ray from LM05.

{\bf NGC~5236 (M~83)}: SFR from Grimm et al. (2003);   metallicity gradient from spectra of HII regions (Pilyugin et al. 2006);  X-ray from LM05.

{\bf NGC~5238 (Mkn~1479)}: SFR from H$\alpha{}$ (Moustakas \&{} Kennicutt 2006b); $Z$ from spectra of HII regions (Moustakas \&{} Kennicutt 2006a);  X-ray from LB05.

{\bf NGC~5408}: SFR from far-infrared (Stevens, Forbes \&{} Norris 2002), adopting the correlation between SFR and FIR in K98; $Z$ from spectra of HII regions (Masegosa, Moles \&{} Campos-Aguilar 1994);  X-ray from LM05. This galaxy is not in the LB05 catalogue. Distance from NED (mean redshift-independent distance). Angular size and axis ratio from the RC3 catalogue; $N_{\rm H}$ from DL90.

{\bf NGC~5457 (M~101, Arp~26)}: average SFR from ultra-violet (SFR$_{{\rm UV}}=5.05\,{}{\rm M}_{\odot}\,{}{\rm yr}^{-1}$, Bell \&{} Kennicutt 2001; Mu$\tilde{\rm n}$oz-Mateos et al. 2007), H$\alpha{}$ (SFR$_{{\rm H}_\alpha{}}=2.58\,{}{\rm M}_{\odot}\,{}{\rm yr}^{-1}$, Kennicutt et al. 1994; Bell \&{} Kennicutt 2001) and radio (SFR$_{1.4{\rm GHz}}=2.02\,{}{\rm M}_{\odot}\,{}{\rm yr}^{-1}$, Condon et al. 2002; Bell 2003; Ranalli et al. 2003) ; metallicity gradient from spectra of HII regions (Pilyugin \&{} Thuan 2007); X-ray from LM05.

{\bf Circinus (PGC~50779)}: SFR from Grimm et al. (2003); estimate of $Z$ from optical lines (Oliva, Marconi \& Moorwood 1999), in agreement with an X-ray estimate of $Z$ (Smith \&{} Wilson 2001). The measurement of $Z$ might be affected by the central AGN; X-ray from LM05. This galaxy is not in the LB05 catalogue. Distance from NED (mean redshift-independent distance). Angular size and axis ratio from the RC3 catalogue; $N_{\rm H}$ from DL90.

{\bf NGC~6946 (Arp~29)}: SFR from radio (Bell 2003); metallicity gradient from spectra of HII regions (Pilyugin et al. 2002); X-ray from LM05.

{\bf PGC~68618 (IC~5201)}: SFR from H$\alpha{}$ (Dopita \&{} Ryder 1994; Ryder \&{} Dopita 1994), adopting the relation in  K98;  metallicity gradient from spectra of HII regions (PVC04);  X-ray from LB05. 

{\bf NGC~7714/7715 (Arp~284, Mkn~538)}: SFR from far-infrared (Schmitt et al. 2006), adopting the relation in  K98; $Z$ from spectra of HII regions (Gonzalez-Delgado et al. 1995); X-ray from Smith, Struck \&{} Nowak (2005).

{\bf NGC~7742}: average SFR from ultra-violet (SFR$_{{\rm UV}}=1.00\,{}{\rm M}_{\odot}\,{}{\rm yr}^{-1}$, Iglesias-P\'aramo et al. 2006), H$\alpha{}$ (SFR$_{{\rm H}_\alpha{}}=1.40\,{}{\rm M}_{\odot}\,{}{\rm yr}^{-1}$, Trinchieri, Fabbiano \&{} Bandiera 1989) and FIR (SFR$_{{\rm FIR}}=1.42\,{}{\rm M}_{\odot}\,{}{\rm yr}^{-1}$, Trinchieri et al. 1989); $Z$ from spectra of HII regions (PT05); X-ray from LB05.

{\bf Milky Way (MW)}: SFR from Grimm et al. (2003); metallicity gradient ($Z(r)=8.762 - 0.356 r/R_{25}$) from OB stars (Daflon \&{} Cunha 2004); X-ray from Grimm, Gilfanov \&{} Sunyaev (2002). This galaxy is not in the LB05 catalogue. Suggested IAU distance to MW centre. Angular size and $N_{\rm H}$ were not used.

{\bf IC~10}: SFR reported in table~2 of Legrand et al. (2001); $Z$ from spectra of HII regions (Garnett 1990), $Z$ from spectra of HII regions; X-ray from Wang, Whitaker \& Williams (2005). This galaxy is not in the LB05 catalogue. Distance from NED (mean redshift-independent distance). Angular size and axis ratio from the RC3 catalogue; $N_{\rm H}$ from DL90.

{\bf Large Magellanic Cloud (LMC)}: SFR from Grimm et al. (2003); metallicity gradient from spectra of HII regions (PVC04); X-ray from Long, Helfand \&{} Grabelsky (1981). This galaxy is not in the LB05 catalogue. Distance from NED (mean redshift-independent distance). Angular size and $N_{\rm H}$ were not used.
 
{\bf Small Magellanic Cloud (SMC)}: SFR from Grimm et al. (2003); metallicity gradient from spectra of HII regions (PVC04); X-ray from Wang \&{} Wu (1992). This galaxy is not in the LB05 catalogue. Distance from NED (mean redshift-independent distance). Angular size and $N_{\rm H}$ were not used.
 

\section{Analysis of $\epsilon{}_{\rm BH}$}
$\epsilon{}_{\rm BH}$ is defined as the ratio between  N$_{\rm ULX}$ and N$_{\rm BH}$ (see equation~\ref{eq:eobs}).
\begin{table*}
\begin{center}
\caption{Parameters of the  power-law fits and  $\chi{}^2$ for $\epsilon{}_{\rm BH}$.} \leavevmode
\begin{tabular}[!h]{lllllll}
\hline
$x-$ axis
& $y-$ axis
& Stellar evolution model
& Index
& Normalization
& $\chi{}^2/{\rm dof}$ $^{\rm a}$
& r (p$_{\rm r}$) $^{\rm b}$\\
\hline
SFR $[{\rm M}_\odot{}\textrm{ yr}^{-1}]$ &   $\epsilon{}_{\rm BH}$ & P98       & $-0.27\pm{}0.19$ &   $-3.02^{+0.09}_{-0.12}$ &  $17.8/62$  & 0.43 ($5\times{}10^{-4}$)\\
 
SFR $[{\rm M}_\odot{}\textrm{ yr}^{-1}]$ &   $\epsilon{}_{\rm BH}$ & P98       & $0.00$ &   $-3.18\pm{}0.07$ &   $19.7/63$ & 0.43 ($5\times{}10^{-4}$)\\

$Z/Z_\odot{}$  &   $\epsilon{}_{\rm BH}$ & P98       & $0.01^{+0.36}_{-0.27}$ &   $-3.17^{+0.27}_{-0.23}$ &   $19.7/62$ & 0.25 ($5\times{}10^{-2}$)\\
 
$Z/Z_\odot{}$  &   $\epsilon{}_{\rm BH}$ & P98       & $0.00$ &   $-3.18\pm{}0.07$ &   $ 19.7/63$ & 0.25 ($5\times{}10^{-2}$)\\

SFR $[{\rm M}_\odot{}\textrm{ yr}^{-1}]$ &   $\epsilon{}_{\rm BH}$ & B10       & $-0.26\pm{}0.19$ &   $-3.27^{+0.10}_{-0.13}$ &   $8.2/62$ & -0.06 ($7\times{}10^{-1}$)\\
 
SFR $[{\rm M}_\odot{}\textrm{ yr}^{-1}]$ &   $\epsilon{}_{\rm BH}$ & B10       & $0.00$ &   $-3.43\pm{}0.06$ &   $10.0/63$ & -0.06 ($7\times{}10^{-1}$)\\

$Z/Z_\odot{}$  &   $\epsilon{}_{\rm BH}$ & B10       & $-0.16\pm{}0.29$ &   $-3.54\pm{}0.22$ &   $9.6/62$ & $-0.23$ ($7\times{}10^{-2}$)\\
 
$Z/Z_\odot{}$  &   $\epsilon{}_{\rm BH}$ & B10       & $0.00$ &   $-3.43\pm{}0.06$ &   $10.0/63$ & $-0.23$ ($7\times{}10^{-2}$)\\

\noalign{\vspace{0.1cm}}
\hline
\end{tabular}
\end{center}
\footnotesize{{\it Notes.} The sample adopted for the fit in this Table is represented by all the galaxies in Table~1.
$^{\rm a}$  $\chi{}^2/{\rm dof}$ is the $\chi{}^2$ divided by the degrees of freedom (dof).
$^{\rm b}$ r is the Pearson correlation coefficient, and p$_{\rm r}$ is the probability of finding a value larger than $|{\rm r}|$, if the two variables were uncorrelated and normally distributed.
}
\end{table*}
\begin{figure*}
\center{{
\epsfig{figure=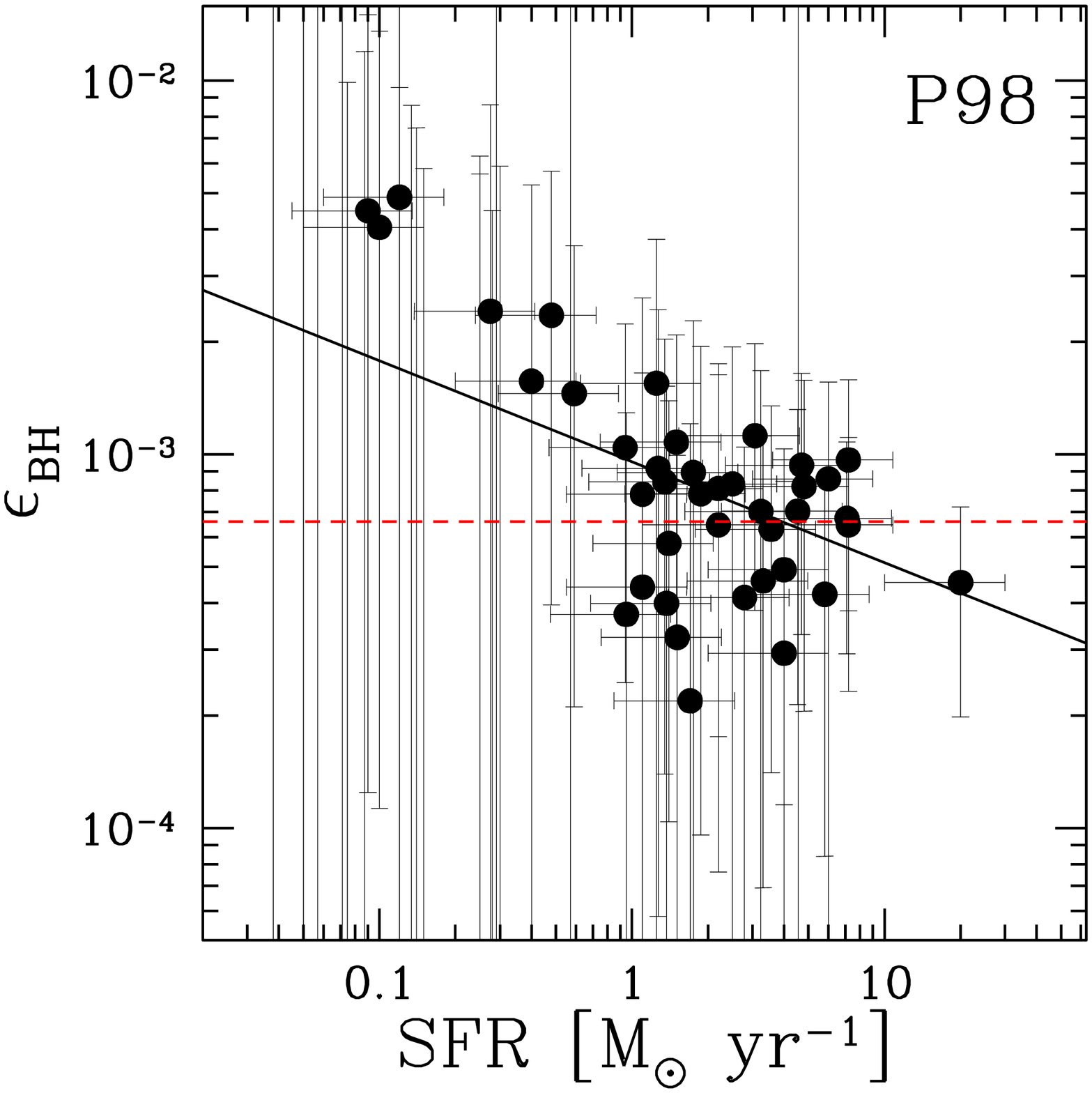,height=7cm} 
\epsfig{figure=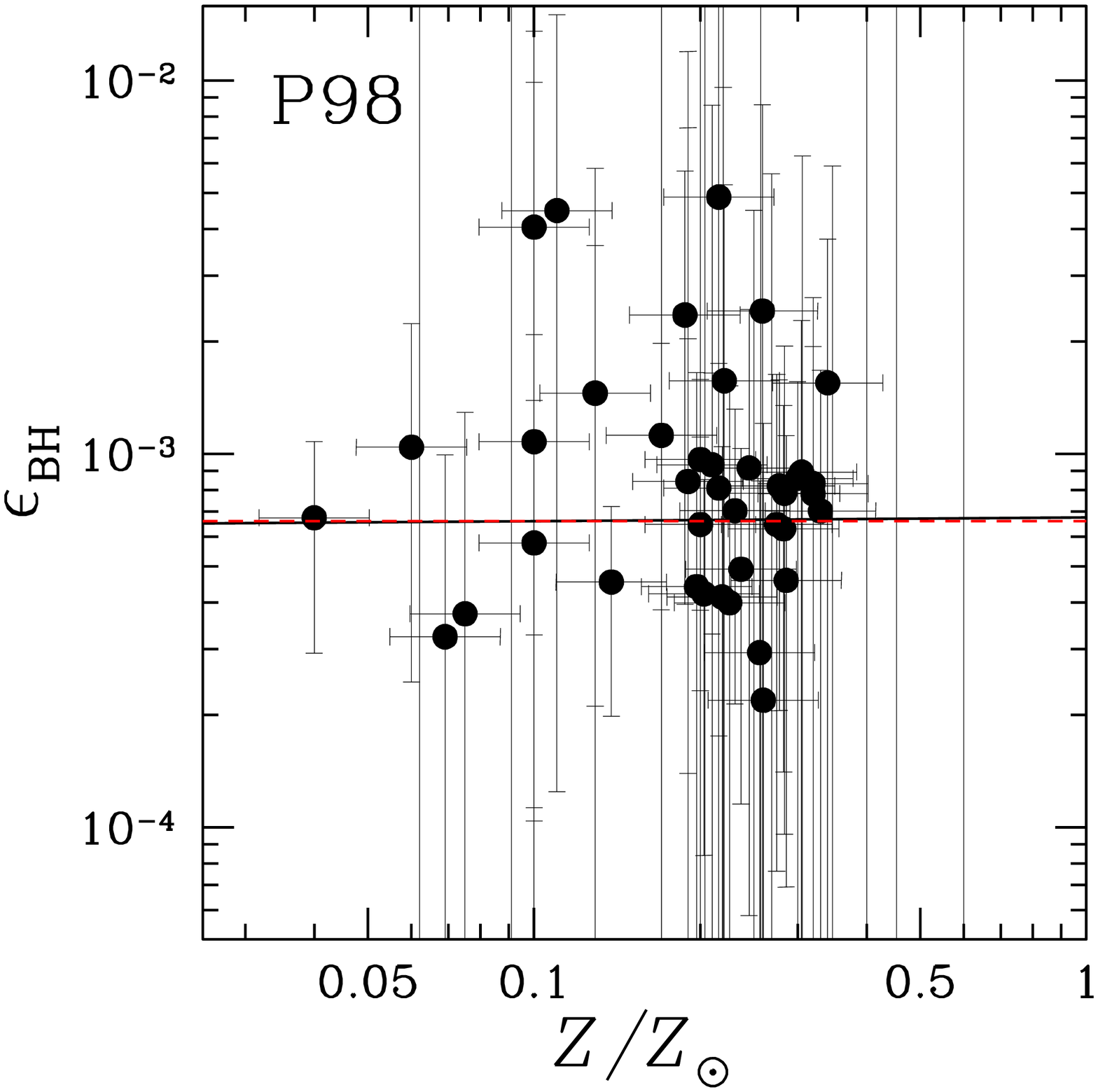,height=7cm} 
}}
\caption{\label{fig:figepsi}
$\epsilon{}_{\rm BH}$ as a function of the SFR (left-hand panel) and of $Z$ (right-hand panel) for P98. The filled circles are the galaxies listed in Table~1. The error bars on both the $x-$ and the $y-$ axis are $1-\sigma{}$ errors (see Section 2.3 for details). The solid lines are the  power-law fits for the sample with N$_{\rm ULX}>0$. The dashed lines (red on the web)  are fits with a constant value for the same sample.
}
\end{figure*}
\begin{figure*}
\center{{
\epsfig{figure=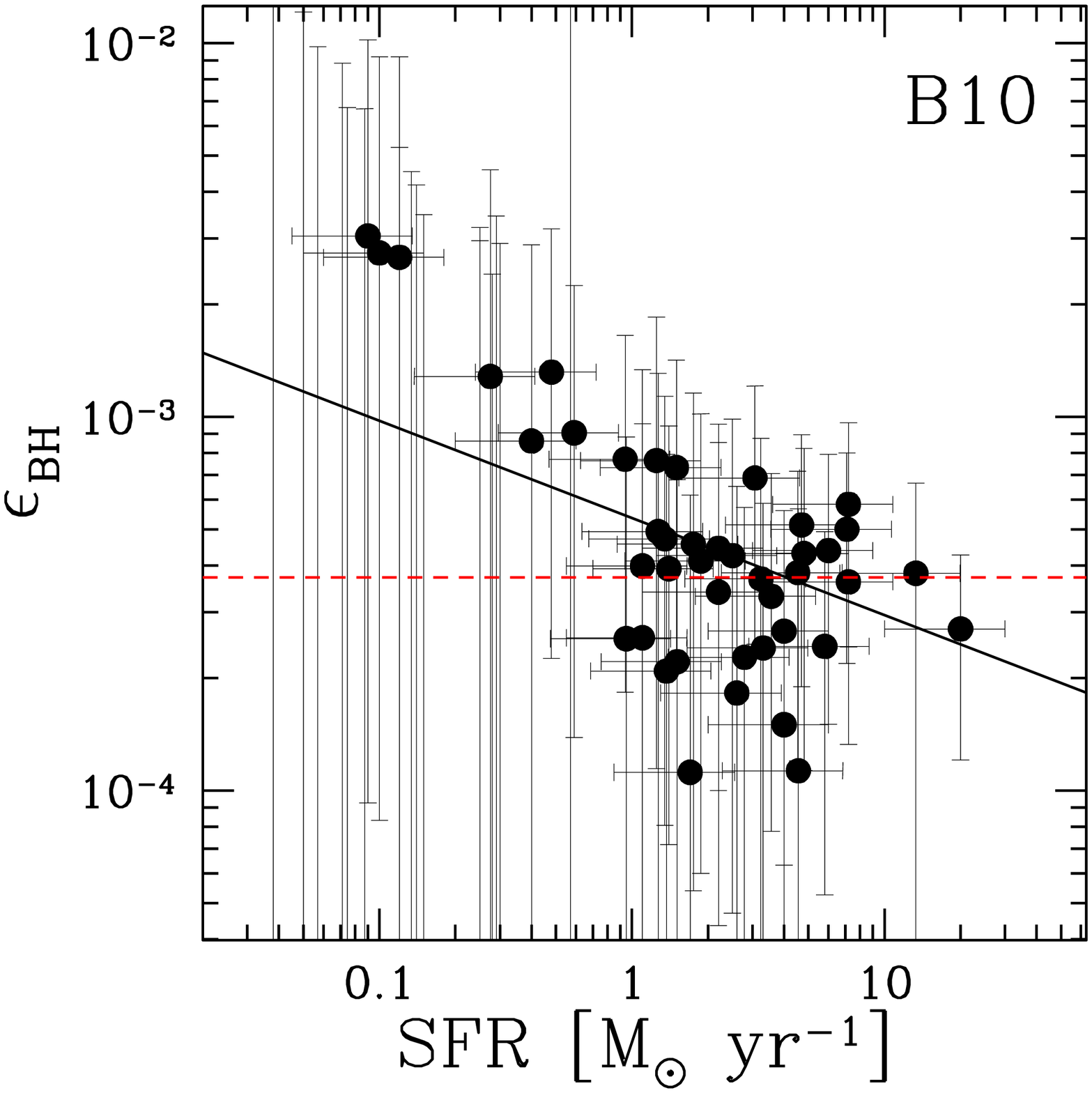,height=7cm} 
\epsfig{figure=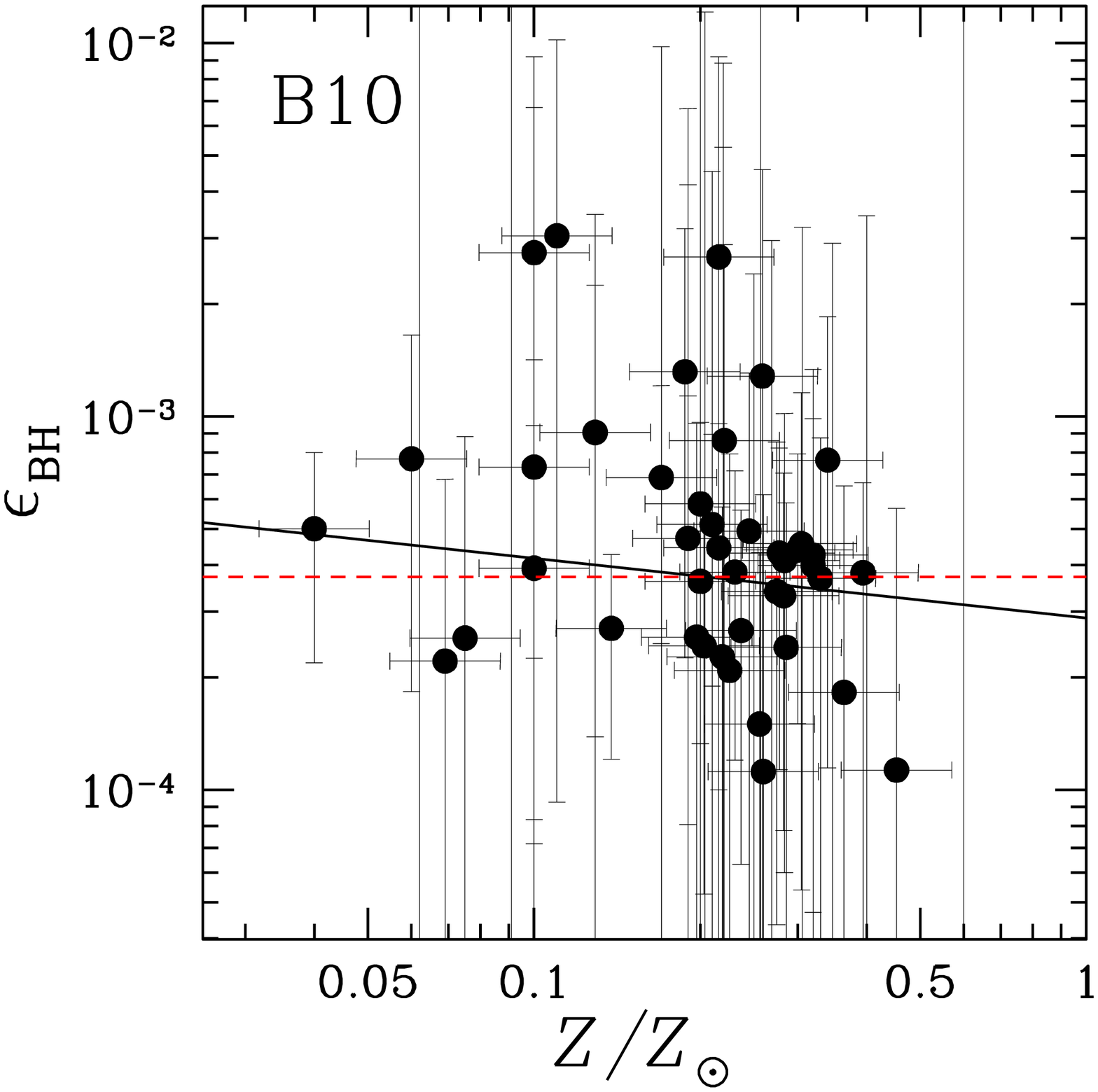,height=7cm} 
}}
\caption{\label{fig:figepsi2}
The same as Fig.~\ref{fig:figepsi} for B10.
}
\end{figure*}
 Table~C1 and Figs.~\ref{fig:figepsi} and \ref{fig:figepsi2} show that the behaviour of $\epsilon{}_{\rm BH}$ as a function of the SFR and of the metallicity is consistent with a constant. This result supports the hypothesis that N$_{\rm BH}$ is proportional to N$_{\rm ULX}$, although the error bars on  $\epsilon{}_{\rm BH}$ are very large.

We point out that the trend that can be seen in the left-hand panels of Figs.~\ref{fig:figepsi} and \ref{fig:figepsi2} ($\epsilon{}_{\rm BH}$ versus SFR) is much less significant than it appears. In fact, in galaxies with a low SFR the presence of a single ULX can easily increase the value of $\epsilon{}_{\rm BH}$, populating the top-left part of the diagrams. We checked the importance of this effect by looking at the 23 galaxies with SFR $\le0.3\,{\rm M}_\odot\,{\rm yr}^{-1}$:
 the average of their $\epsilon{}_{\rm BH}$, weighted on N$_{\rm BH}$, is $\sim(9\pm5)\times10^{-4}$, in the P98 case. Such value is much more in line with the results for galaxies with higher SFR; it is still a bit higher than the average, but this might be due to some form of publication/observation bias.

\end{document}